\documentclass[iop]{emulateapj}
\usepackage{epsfig}
\usepackage{apjfonts}
\usepackage{color}
\usepackage{aas_macros}
\usepackage{bm}
\usepackage{subfigure}
\usepackage{textcomp}

\begin{document}

\title{A further study of $t_{\rm burst}$ of GRBs: rest frame properties, external plateau contributions and multiple parameter analysis}
\author{He Gao$^{1,2*}$, An-Bing Ren$^1$, Wei-Hua Lei$^3$, Bin-Bin Zhang$^{4,5}$, Hou-Jun L\"{u}$^{2,6}$ and Ye Li$^7$}

\affiliation{
$^1$Department of Astronomy, Beijing Normal University, Beijing 100875, China; gaohe@bnu.edu.cn\\
$^2$Guangxi Key Laboratory for Relativistic Astrophysics, Nanning, Guangxi 530004, China;\\
$^3$School of Physics, Huazhong University of Science and Technology, Wuhan, 430074, China;\\
$^4$Instituto de Astrof\'isica de Andaluc\'a (IAA-CSIC), P.O. Box 03004, E-18080 Granada, Spain; \\ 
$^5$Scientist Support LLC, Madsion, AL 35758, USA;\\ 
$^6$GXU-NAOC Center for Astrophysics and Space Sciences, Department of Physics, Guangxi University, Nanning 530004, China;\\
$^7$Department of Physics and Astronomy, University of Nevada Las Vegas, NV 89154, USA.}

\begin{abstract}

\cite{zhang14} propose to redefine the true GRB central engine activity duration, $t_{\rm burst}$, by considering the contributions from the prompt $\gamma$-ray emission, X-ray flare and internal plateau features. 
With a comprehensive study on a large sample of {\em Swift} GRBs, it is shown that the $t_{\rm burst}$ distribution in the observer frame consists of a bimodal feature, suggesting the existence of a new population of ultra-long GRBs.
In this work, we make a series of further studies on $t_{\rm burst}$: we update the {\em Swift} GRB sample up to June 2016; we investigate the properties of $t_{\rm burst}$ distribution in the rest frame; we redefine $t_{\rm burst}$ by involving external plateau contributions; we make a multiple parameter analysis to investigate whether the bursts within the ultra-long population being statistically different in sense of other features besides the duration distribution. We find that for all situations, the distribution of $t_{\rm burst}$ requires two normal distributions in logarithmic space to 
provide a good fit, both in observer frame and rest frame. Considering the observational gap effect would not completely erase the bimodal distribution feature. However the bursts within the ultra-long population may have no statistically different in sense of other features besides the duration term. We thus suggest that if the ultra-long population of GRBs indeed exists, their central engine mechanism and radiation mechanism should be similar to the normal population, but they have longer central engine activity timescale.

\end{abstract}


\section{Introduction}

Gamma-Ray Bursts (GRBs) are the most extreme explosive events in the universe \cite[][for a recent review]{kumarzhang15}. Based on their temporal and spectral statistical properties, GRBs were classified into two categories: long-duration, soft-spectrum class (LGRBs) and the short-duration, hard-spectrum class (SGRBs) \citep{kouveliotou93}. After decades of observations, it turns out that such a phenomenological classification indeed imply different natures, e.g., different type of progenitor were invoked for these two different class. SGRBs are connected with mergers of two compact stellar objects (NS-NS and NS-BH systems) \citep{paczynski86,eichler89,paczynski91,narayan92} and LGRBs are connected with core collapse from Wolf-Rayet star \citep{woosley93,paczynski98,macfadyen99,woosley06}. 

Typically, the prompt duration of LGRBs is tens of seconds. However, there is a subclass of LGRBs  (e.g. GRBs 101225A, 111209A, 121027A and 130925A) showing unusually long prompt duration, as long as hours \citep{gendre13,virgili13,stratta13,levan14,greiner15}. In these references, GRBs with $\gamma$-ray duration $T_{90}$ comparable or larger than $10^3$ s were defined as ``ultra-long GRBs (ulGRBs)". Some authors proposed that these ulGRBs may belong to a new population \citep{gendre13,nakauchi13,levan14,greiner15,ioka16}, and they may either issue from a new type of progenitor, such as blue supergiants \citep{meszaros01,nakauchi13} or dwarf tidal disruption events \citep{ioka16}, or they may have a special central engine, such as a strongly magnetized millisecond neutron star (a magnetar) \citep{levan14,greiner15}. 

\cite{virgili13} investigates the $\gamma$-ray duration distribution of LGRBs, and they claim that 
the overall distribution is consistent with a log-normal distribution, namely ulGRBs are the tail of the distribution of normal LGRBs rather than corresponding to a new possible population. 

However, many Swift GRBs shows interesting features in their X-ray light curves, such as flares \citep{burrows05,zhang06,margutti11} and shallow decay plateaus \citep{troja07,liang07}, signifying an extended central engine activity time. So that it has been widely argued that the prompt duration may not be able to reflect the intrinsic central engine activity. Some authors propose to redefine the burst duration (e.g. $t_{\rm burst}$) by taking into account both $\gamma$-ray and the aforementioned X-ray light curve features \citep{zhang14,boer15}. Such a definition is not easy to quantify, since late time X-ray features need not necessarily be related to late central engine activity. The observed ($\gamma$-ray and X-ray) flux is contributed by both internal dissipation emission (e.g. internal shocks or magnetic dissipation), and the afterglow emission from the external shock. The prompt $\gamma$-ray emission, X-ray flares and the so-called ``internal X-ray plateau" (a plateau in the light curve followed by a very rapid decay) likely originate from internal dissipation, which essentially reflect the intrinsic central engine activity \citep{zhang06,nousek06}. However, the so-called ``external X-ray plateau" (a plateau in the light curve followed by a normal decay as expected from the external shock model) likely originate from external shock emission, and the plateau phase might be due to the late central engine energy injection, but it could also be due to the internal collisions or refreshed external collisions from early ejected shells \citep{reesmeszaros98,sari00,gao13}. For the latter case, the external plateau phase no longer reflects the intrinsic central engine activity. 

For a secure lower limit, $t_{\rm burst}$ could be defined by the last steep-to-shallow transition in the observed ($\gamma$-ray and X-ray) flux, which essentially incorporates the prompt $\gamma$-ray emission, X-ray flares and internal plateau phase \citep{zhang14,boer15}. With a comprehensive study on a large sample of Swift GRBs, it is shown that the engine activity time is frequently much larger than $T_{90}$ \citep{zhang14}. Even the $T_{90}$ distribution could be well fit by a log-normal distribution, the $t_{\rm burst}$ distribution consists of a much larger tail, that requires an additional component to provide a good fit \citep{zhang14,boer15}. However, due to the low significance, \cite{zhang14} and \cite{boer15} have different opinions over the interpretation of this tail, with \cite{zhang14} suggesting that the bimodal distribution of $t_{\rm burst}$ may be strongly affected by some selection effects, so that 
an ultra-long population cannot be confirmed, while \cite{boer15} infer that the ultra-long population is statistically different. 

Recently,  within the framework of the internal-external shock model, \cite{gao15} develop a numerical code to study the relationship between $T_{90}$, $t_{\rm burst}$ as well as the intrinsic central engine activity timescale $T_{\rm ce}$. They found that the values of $T_{\rm 90}$ and $t_{\rm burst}$ could be larger than $T_{\rm ce}$ due to internal collisions or refreshed external collisions from early ejected shells, but this is only valid when $T_{\rm ce}\lesssim 10^4$ s. In other words, ``external X-ray plateau" could also reflect the intrinsic central engine activity as long as $T_{\rm ce}\gtrsim 10^4$ s. 

In this work, we systematically investigate Swift GRBs from the launch of Swift to June 2016. We attempt to answer the following interesting questions: 1) both analysis of \cite{zhang14} and \cite{boer15} focus on the duration distribution in the observed frame. Whether the conclusion becomes different when considering the duration distribution in the rest frame? 2) whether the distribution tail of $t_{\rm burst}$ would become more significant when ``external X-ray plateau" being invoked? 3) if the bimodal distribution of $t_{\rm burst}$ indeed exists, whether the bursts within the ultra-long population being statistically different in sense of other features besides the duration distribution, for instance, does these two populations shows distinct separation in the multiple parameter analysis, such as in the $E_{\rm p,z}-E_{\rm \gamma, iso}$ and $E_{\rm p,z}-L_{\rm \gamma, iso}$ diagrams? 

\section{Data analysis}

Between January 2005 and June 2016, 1032 Swift GRBs were detected by Swift/XRT, with 728 GRBs having well-sampled XRT light curves, namely the X-ray light curve contains at least 6 data points, excluding upper limits. In order to measure $t_{\rm burst}$, we download the XRT light curves from the Swift/XRT team website\footnote{\url{http://www.swift.ac.uk/xrt\_ curves/}} \citep{evans09} at the UK Swift Science Data Centre (UKSSDC), which were processed with HEASOFT v6.12. We then apply a multivariate adaptive regression splines (MARS) technique \citep[e.g.,][]{friedman91} to the observed light curves in the logarithmic (log) scale. MARS is a non-parametric regression technique that could automatically determine both variable selection and functional form, resulting in an explanatory predictive model. Such MARS model can be expressed as a linear combination of piecewise polynomial basis functions (include constant and the so-called Hinge functions) that are joined together smoothly at the knots. When using the MARS for modeling the relationship between the predictor and dependent variables, it is not necessary to know the functional forms of the relation-ships, MARS establishes them based on the data \cite[][for a detailed overview of MARS technique]{Kweku-Muata Osei-Bryson}. Applying to Swift/XRT data, MARS could automatically fit the light curves with multisegment broken power-law functions, detect and optimize all breaks and record the power-law indices for each segment. The results from such a technique are consistent with the automated light curve fitting results provided by the XRT GRB online catalog (see appendix for more details). Nevertheless, such a technique could also incorporate the steep decay and flare phases, which are essential to measure $t_{\rm burst}$ \citep{zhang14}. 

We then distribute 728 bursts into four categories:  

Bursts in the first three categories do not consist X-ray flares:  (1) 308 bursts either with simple power-law decay light curves, or with broken power-law decay light curves but without showing any steep decay\footnote{For steep decay, the steepest decay slope in an external shock model is $2 + \beta$ \citep{kumar00}, which is typically smaller than 3, and is defined by the high-latitude ``curvature effect" emission from a conical outflow, even if the emission abruptly ceases.} (power-law index steeper than $-3$) or plateau signature\footnote{For plateau phase, there is no stringent definition on its decay slope. However, according to standard external shock model \cite[][for a review]{gao13}, a natural interpretation of the plateau phase is to attribute it to a continuous energy injection, so that the forward shock is ``refreshed". The energy injection process could be effectively interpreted as the central engine itself being longer lasting, with a power-law luminosity history $L\propto t^{-q}$. The effective q value inferred from the observations is around 0.5 \citep{zhang06}. Adopting $q=0.5$ and assuming the electron spectral index $p=2.2$, we have the plateau decay index as $\sim0.6$, which is taken as the criteria of plateau decay slope.} (defined as a temporal segment with decay slope $\lesssim0.6$); (2) 40 bursts with steep decay feature but without showing any flare or plateau feature; (3) 145 bursts without flare feature but consisting plateau feature in the light curve. Finally, 235 bursts consisting X-ray flare feature are distributed into the fourth category. 

For the first category (e.g. 308 GRBs), we collect their gamma-ray duration $T_{90}$, their redshift $z$ (if available), their spectral parameters such as the peak energy in the energy spectrum $E_{p}$, spectral index of power law fitting $\Gamma$, the isotropic gamma-ray energy $E_{\gamma,\rm iso}$ and peak luminosity $L_{\gamma,\rm iso}$. Please refer to \cite{li16} for details of data collecting and spectral parameters definition and calculation. 

For the second category (e.g. 40 GRBs), we record the transition time of steep decay to normal decay (defined as a temporal segment with decay slope $0.6<\alpha<3$ ) as $t_{\rm stp}$.

Within the third category (e.g. 145 GRBs), 10 bursts consist internal plateaus and 135 bursts consist external plateaus. For the internal plateau sample, the plateau is followed by a steep decay segment. We record the end of this steep decay phase as $t_{\rm stp}$. For the external plateau sample, we record the last plateau to normal decay transition time as the ending time of the plateau, $t_{\rm pla}$. Note that among the external plateau sample, we have 62 GRBs with $t_{\rm pla}\gtrsim10^{4}~\rm{s}$, and 73 GRBs with $t_{\rm pla}<10^{4}~\rm{s}$. 

Within the fourth category  (e.g. 235 GRBs), 50 GRBs show additional plateau feature after the last X-ray flare, with 3 being internal plateau and 47 being external plateau. For all 235 bursts, we record the end of the last steep decay phase as $t_{\rm stp}$. For those 185 GRBs without plateau, $t_{\rm stp}$ corresponding to the ending time of the last flare. For those 3 bursts with internal plateau, $t_{\rm stp}$ corresponding to the ending time of the last steep decay phase following the plateau. For those 47 bursts showing external plateaus, we record the last plateau to normal decay transition time as the ending time of the plateau, $t_{\rm pla}$.

For all GRBs in category (2), (3) and (4), we also collect their gamma-ray duration $T_{90}$, redshift $z$  (if available), and their spectral parameters similar to category (1). Note that in the appendix (Figure 5), we show some examples of our fitting results for different types of light curves, with marks of $t_{\rm stp}$ and $t_{\rm pla}$ (if applicable), to better illustrate the light curve properties for each of the four categories. 

\begin{figure*}
\label{fig:1}
    \subfigure[]{
    \includegraphics[width=2.0in]{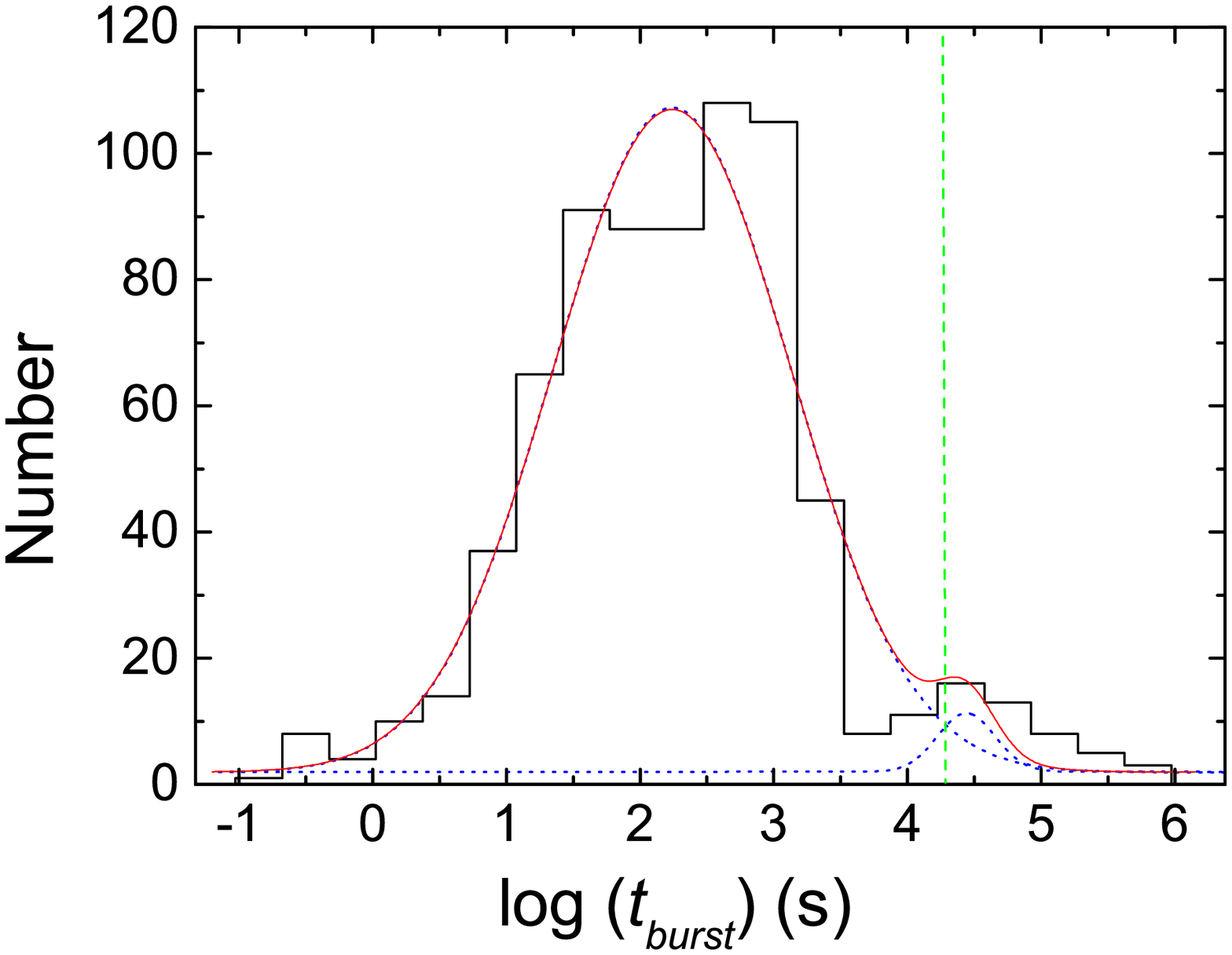}}
    \hspace{-0.4in}
    \subfigure[]{
    \includegraphics[width=2.0in]{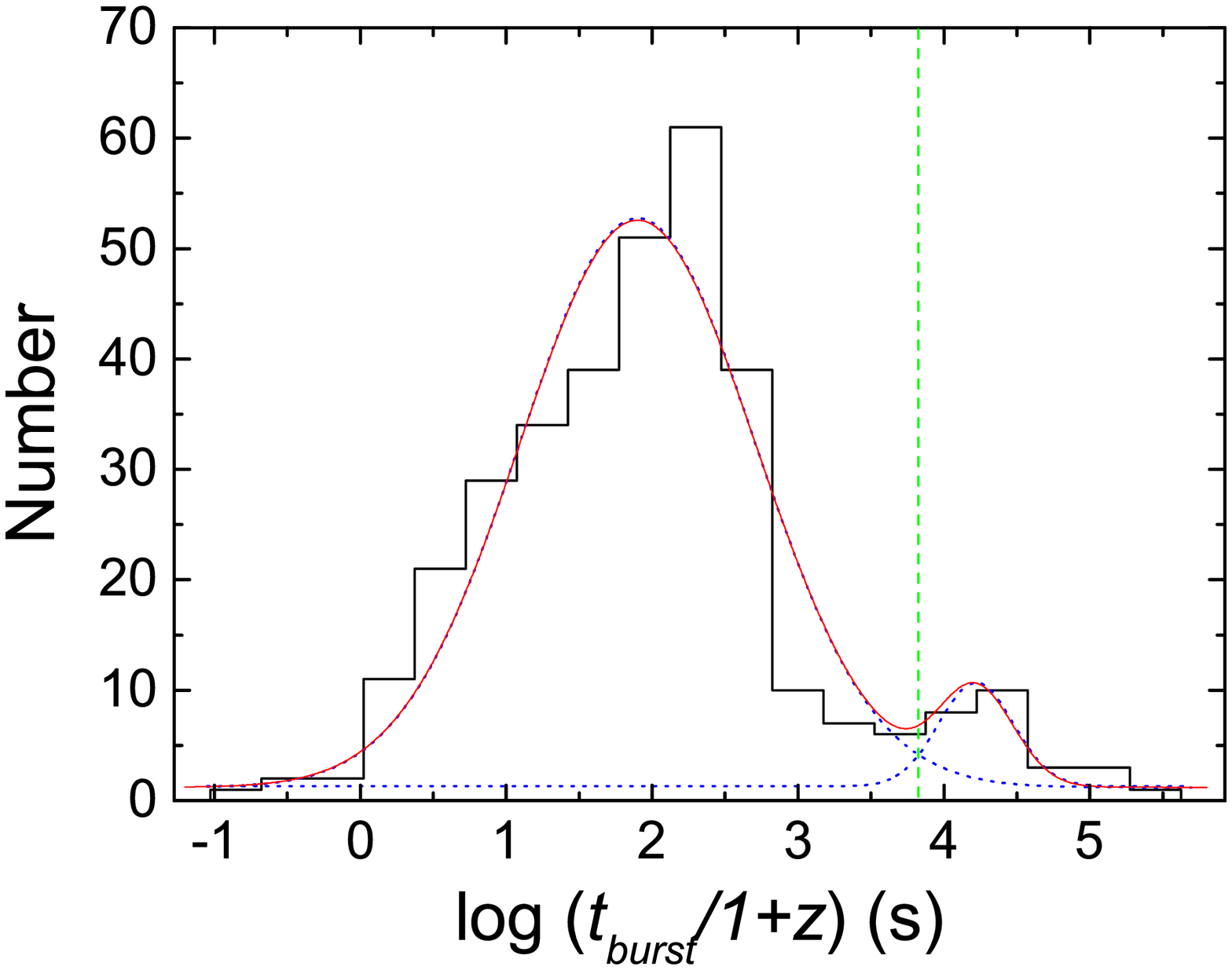}}
    \hspace{-0.4in}
     \subfigure[]{
       \includegraphics[width=2.0in]{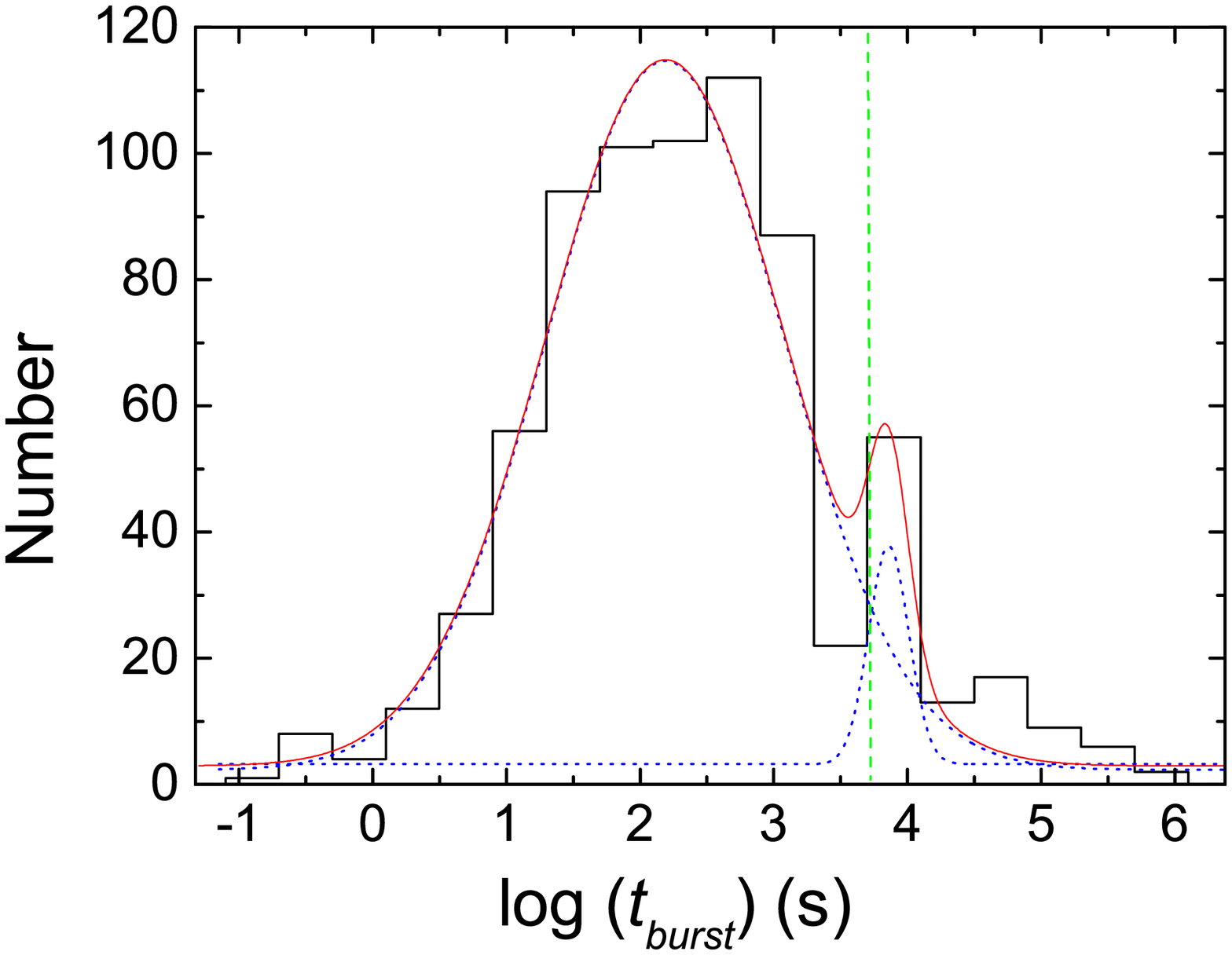}}
       \hspace{-0.4in}
    \subfigure[]{
    \includegraphics[width=2.0in]{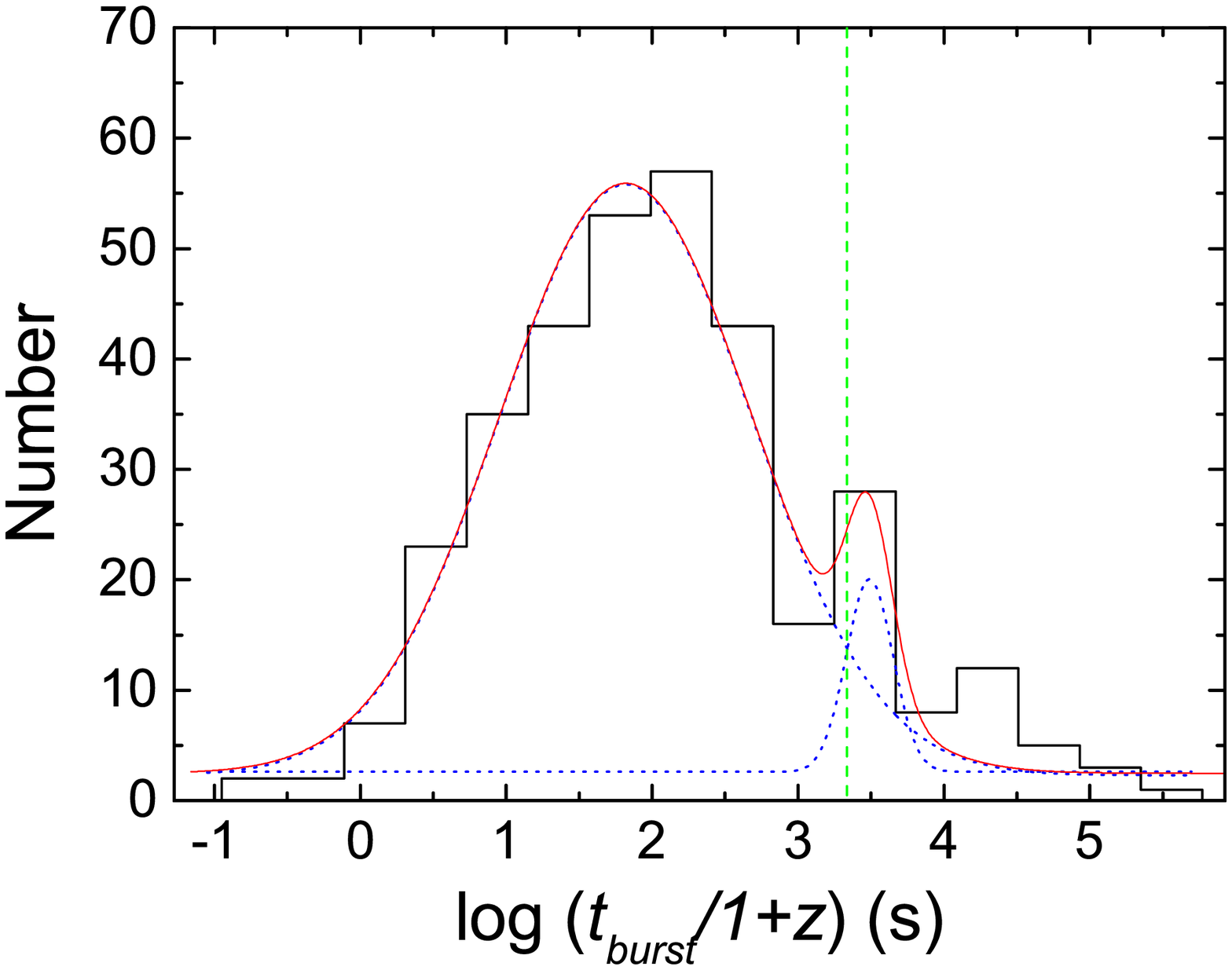}}\\
     \subfigure[]{
    \includegraphics[width=2.0in]{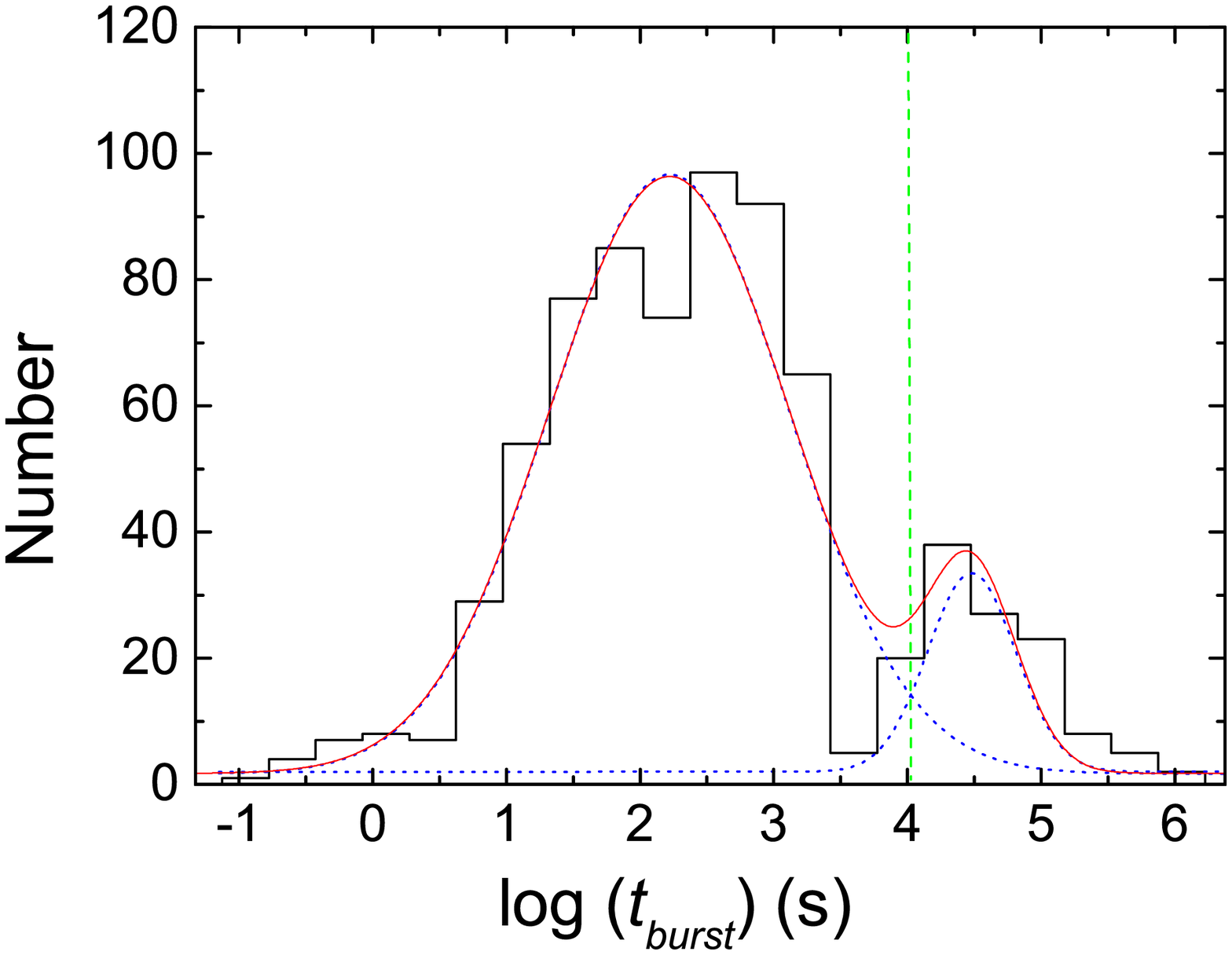}}
    \hspace{-0.4in}
    \subfigure[]{
    \includegraphics[width=2.0in]{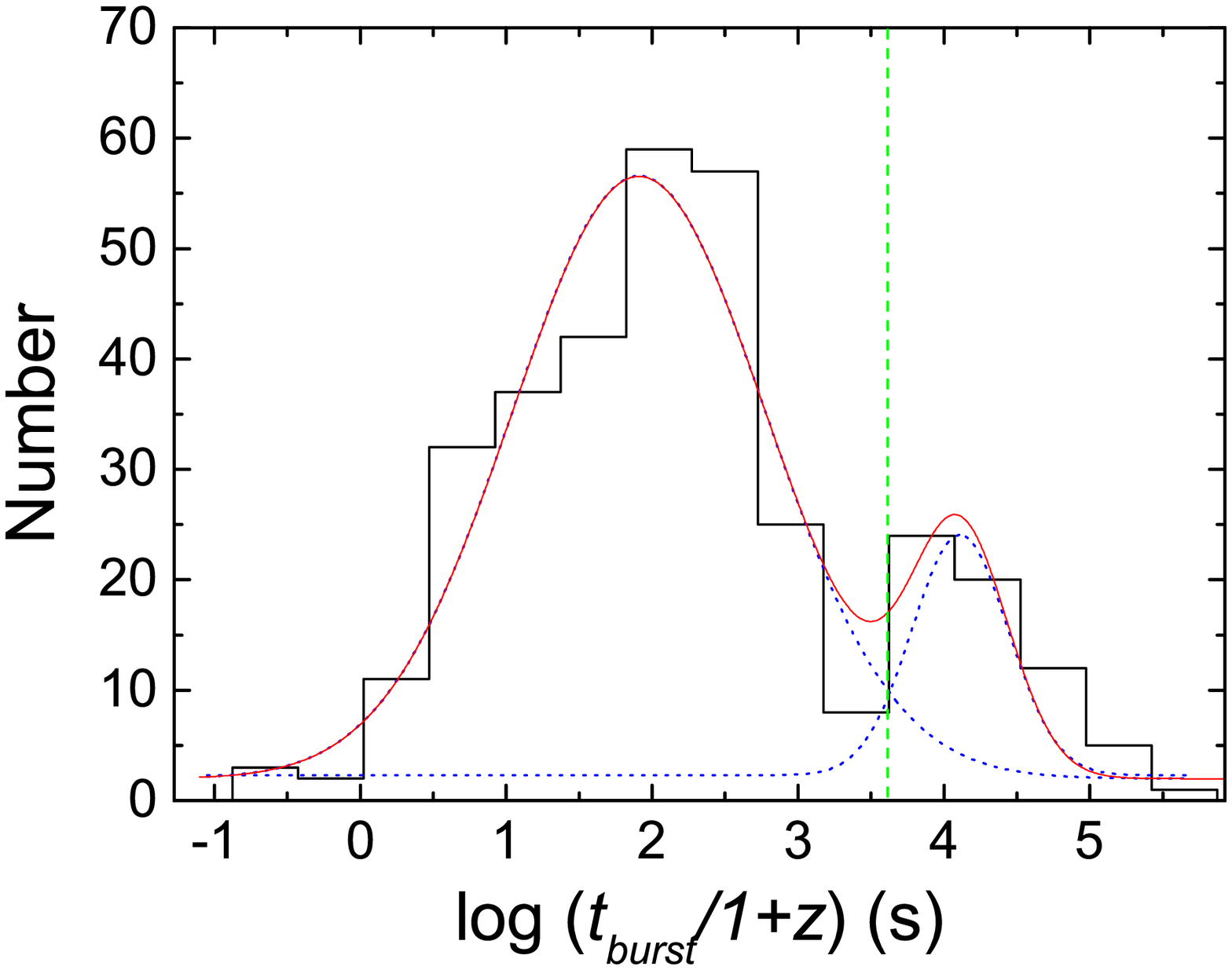}}
    \hspace{-0.4in}
     \subfigure[]{
       \includegraphics[width=2.0in]{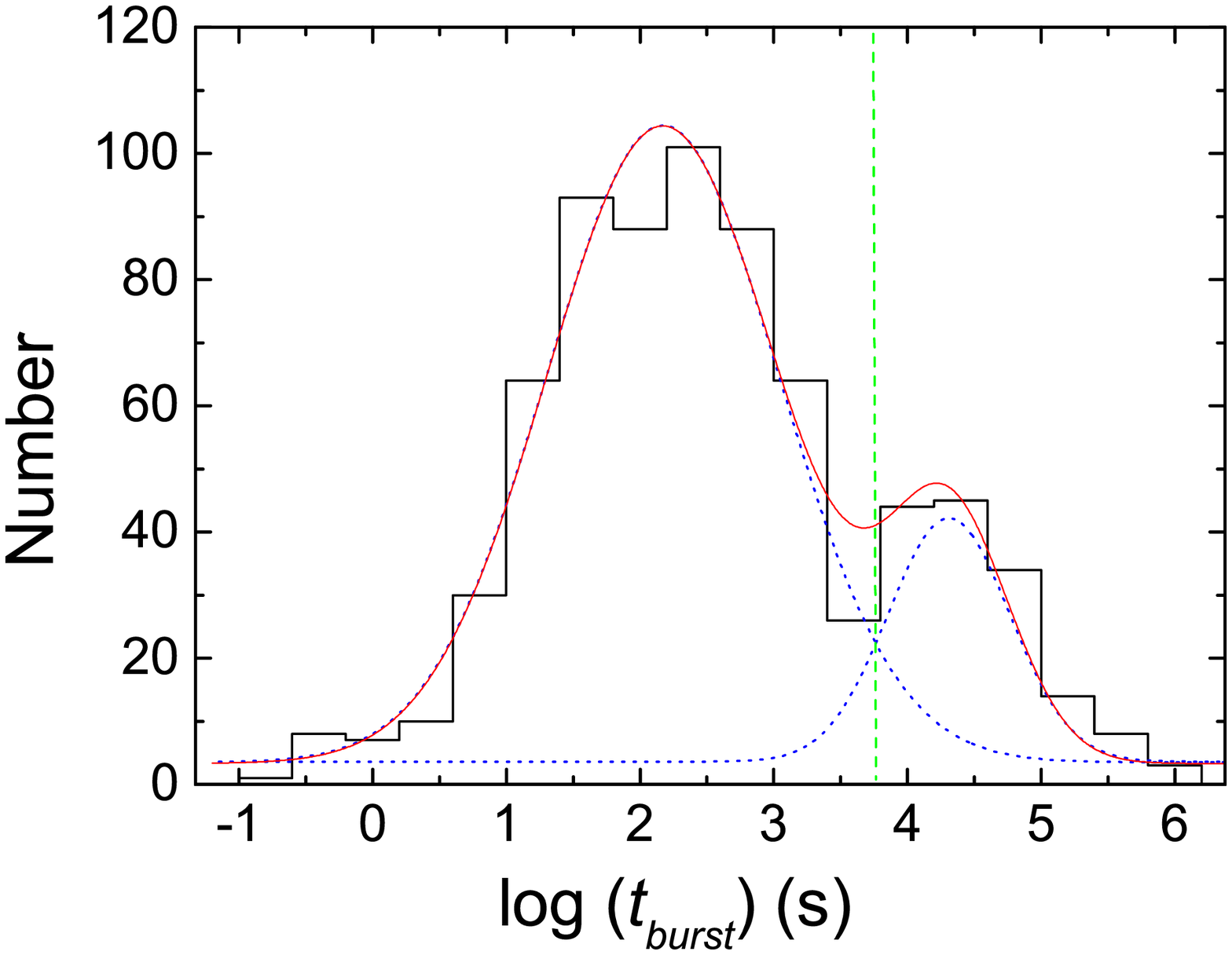}}
       \hspace{-0.4in}
    \subfigure[]{
    \includegraphics[width=2.0in]{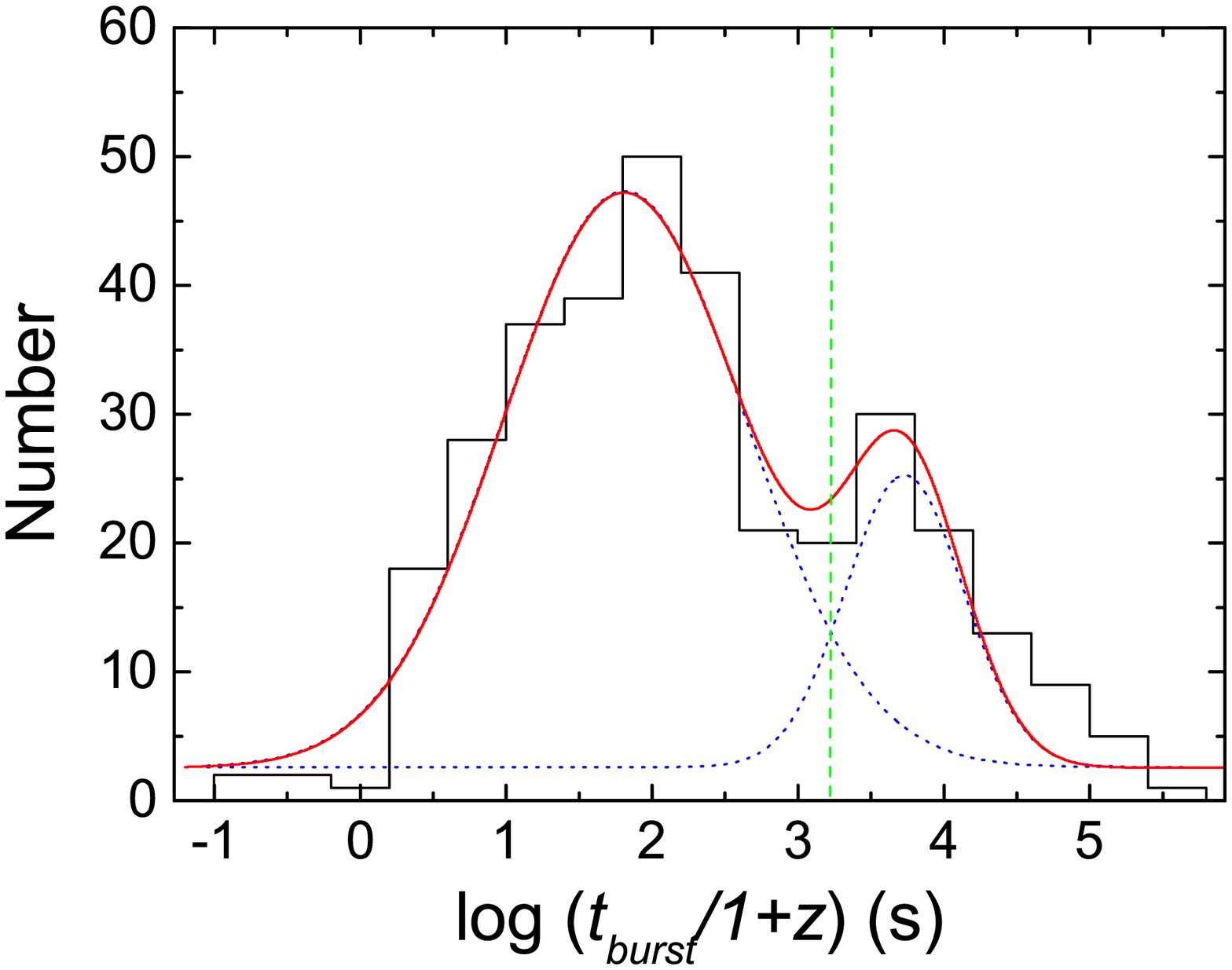}}\\
    \subfigure[]{
    \includegraphics[width=2.0in]{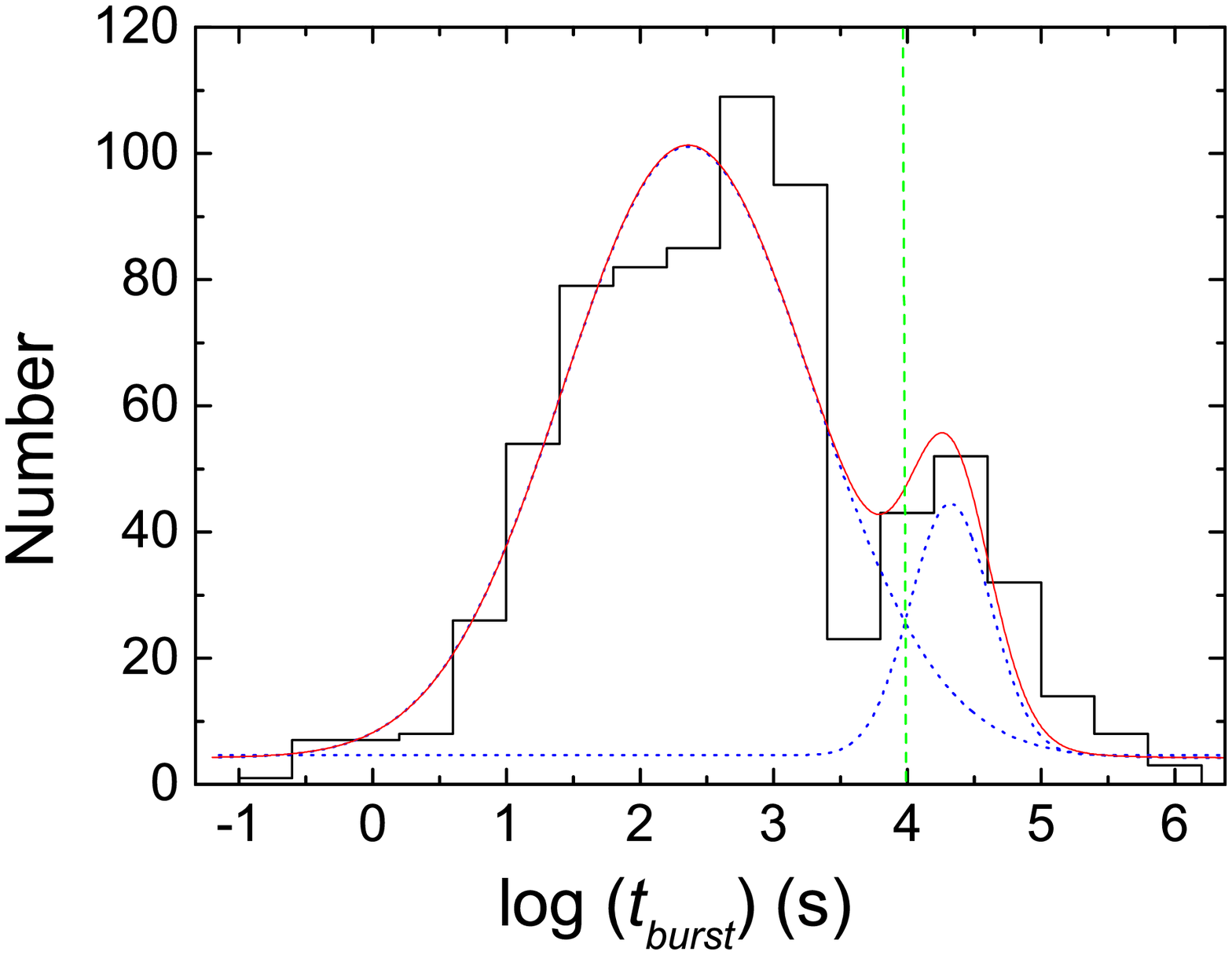}}
    \hspace{-0.4in}
    \subfigure[]{
    \includegraphics[width=2.0in]{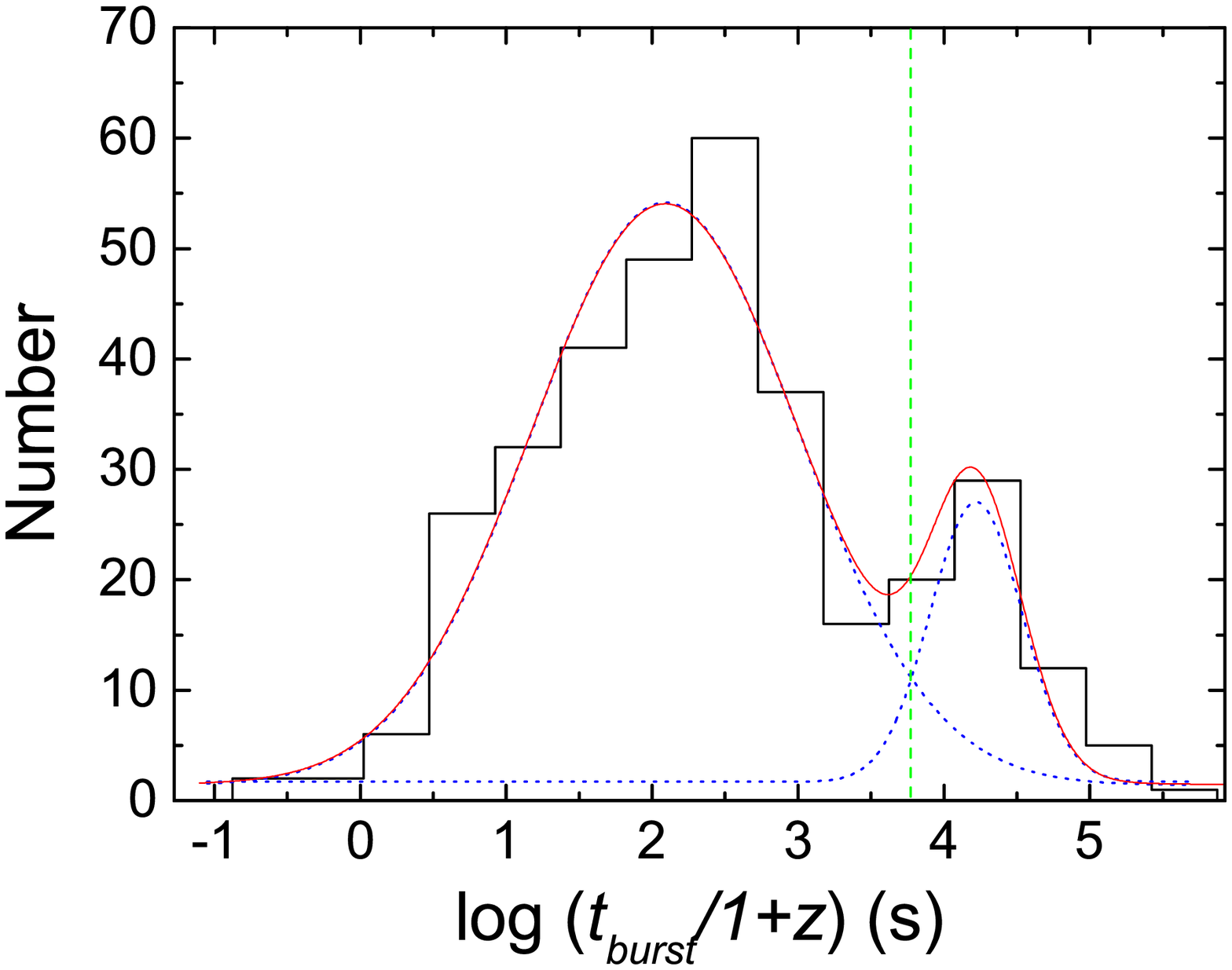}}
    \hspace{-0.4in}
     \subfigure[]{
       \includegraphics[width=2.0in]{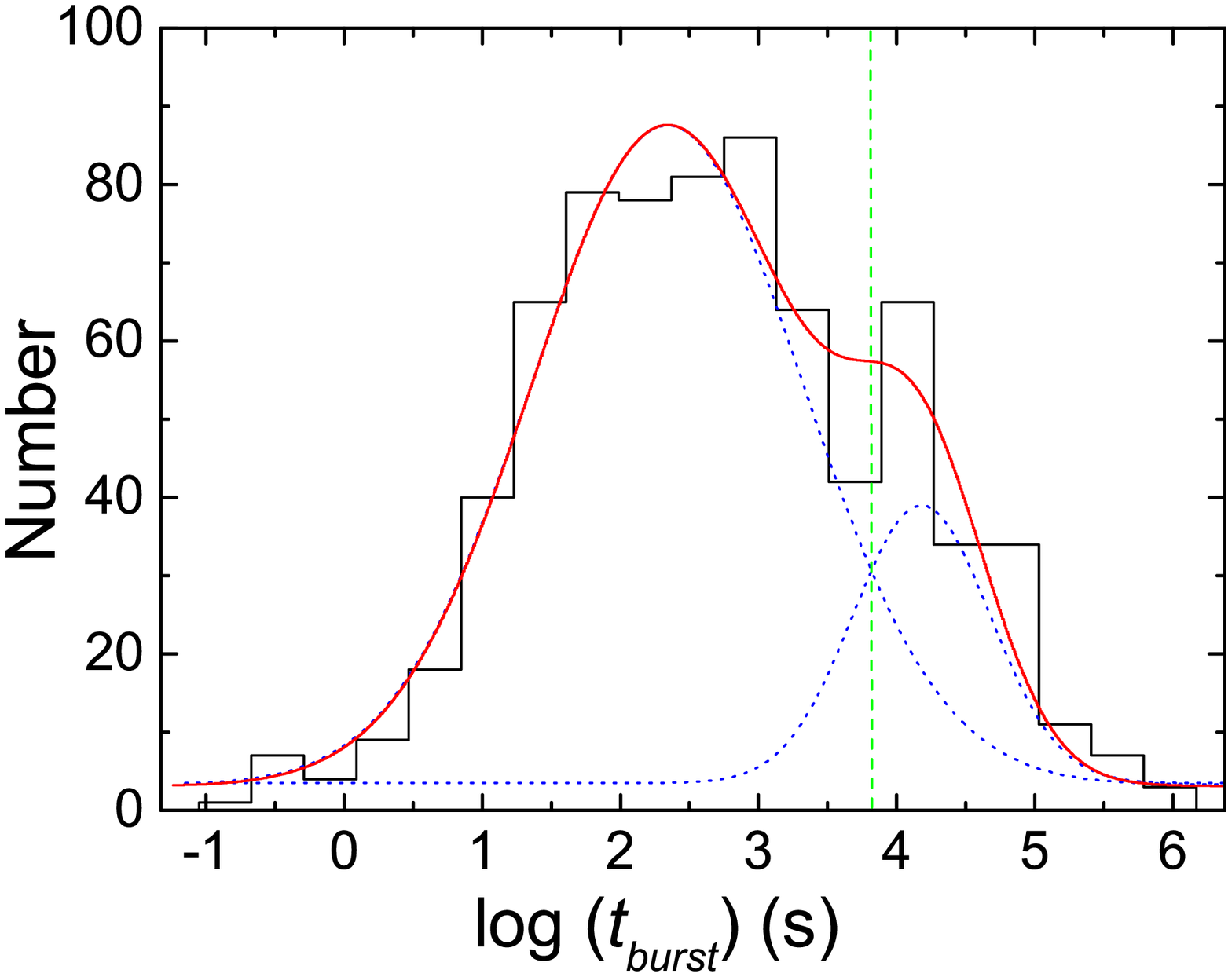}}
       \hspace{-0.4in}
    \subfigure[]{
    \includegraphics[width=2.0in]{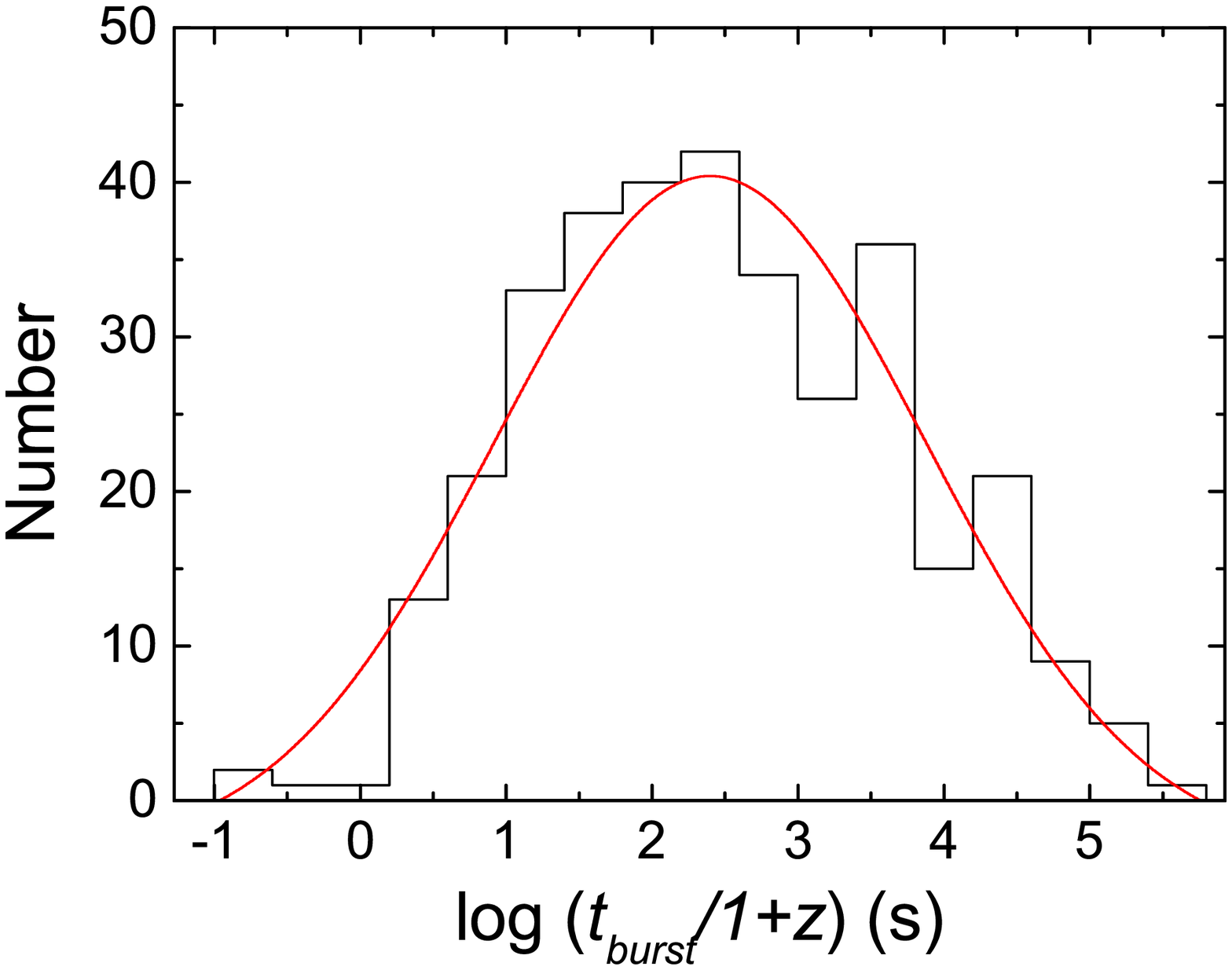}}\\
    \subfigure[]{
    \includegraphics[width=2.0in]{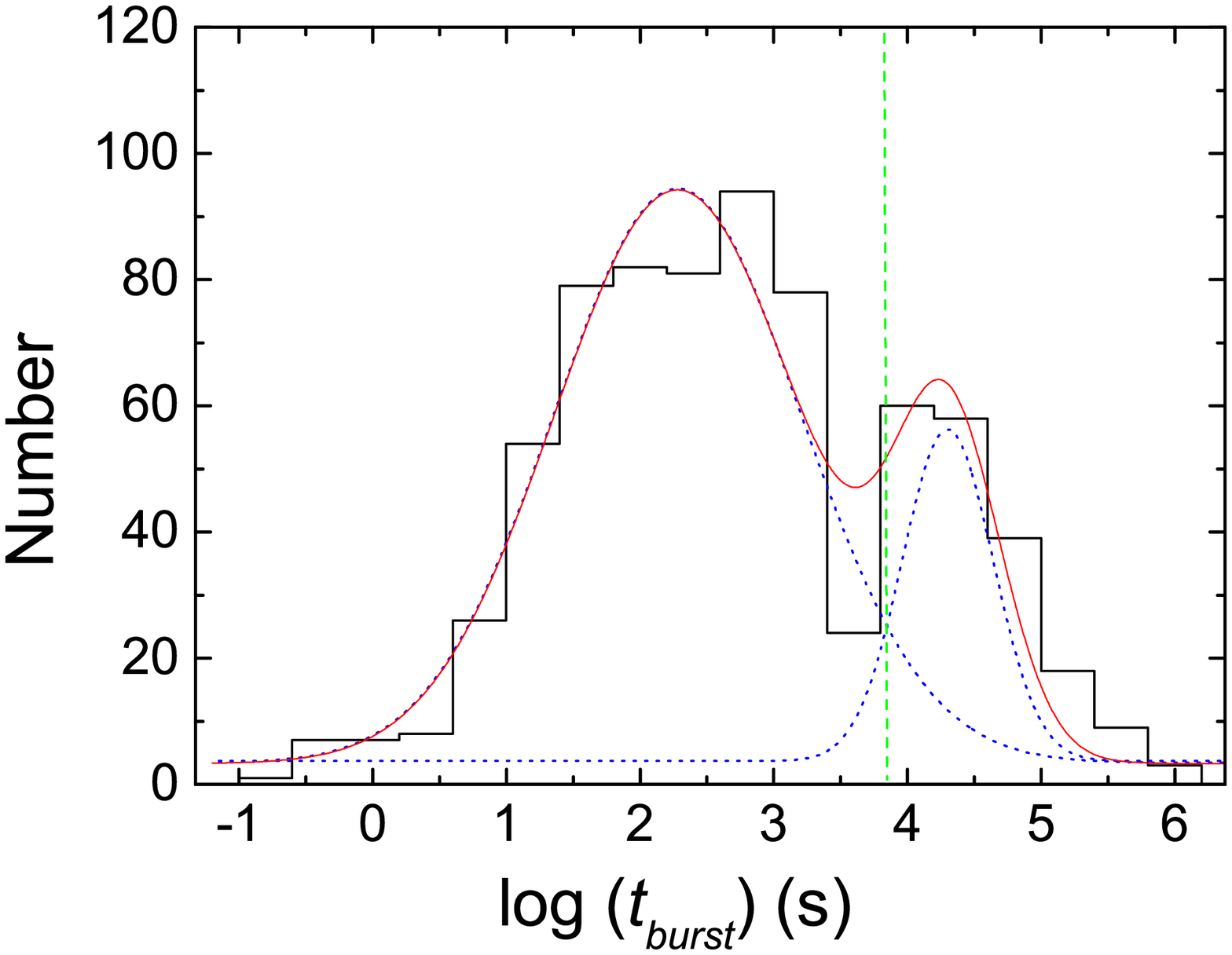}}
    \hspace{-0.4in}
    \subfigure[]{
    \includegraphics[width=2.0in]{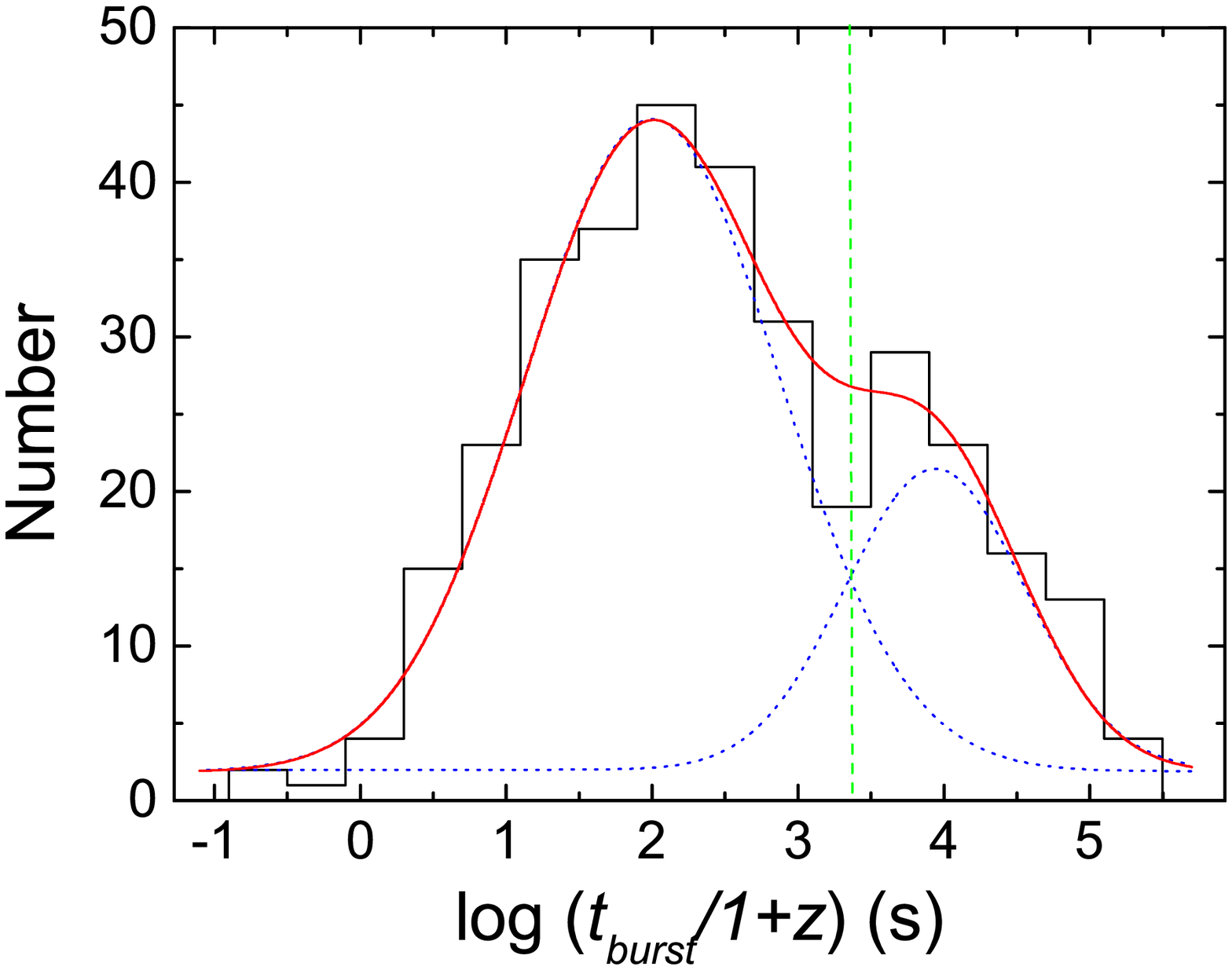}}
    \hspace{-0.4in}
     \subfigure[]{
       \includegraphics[width=2.0in]{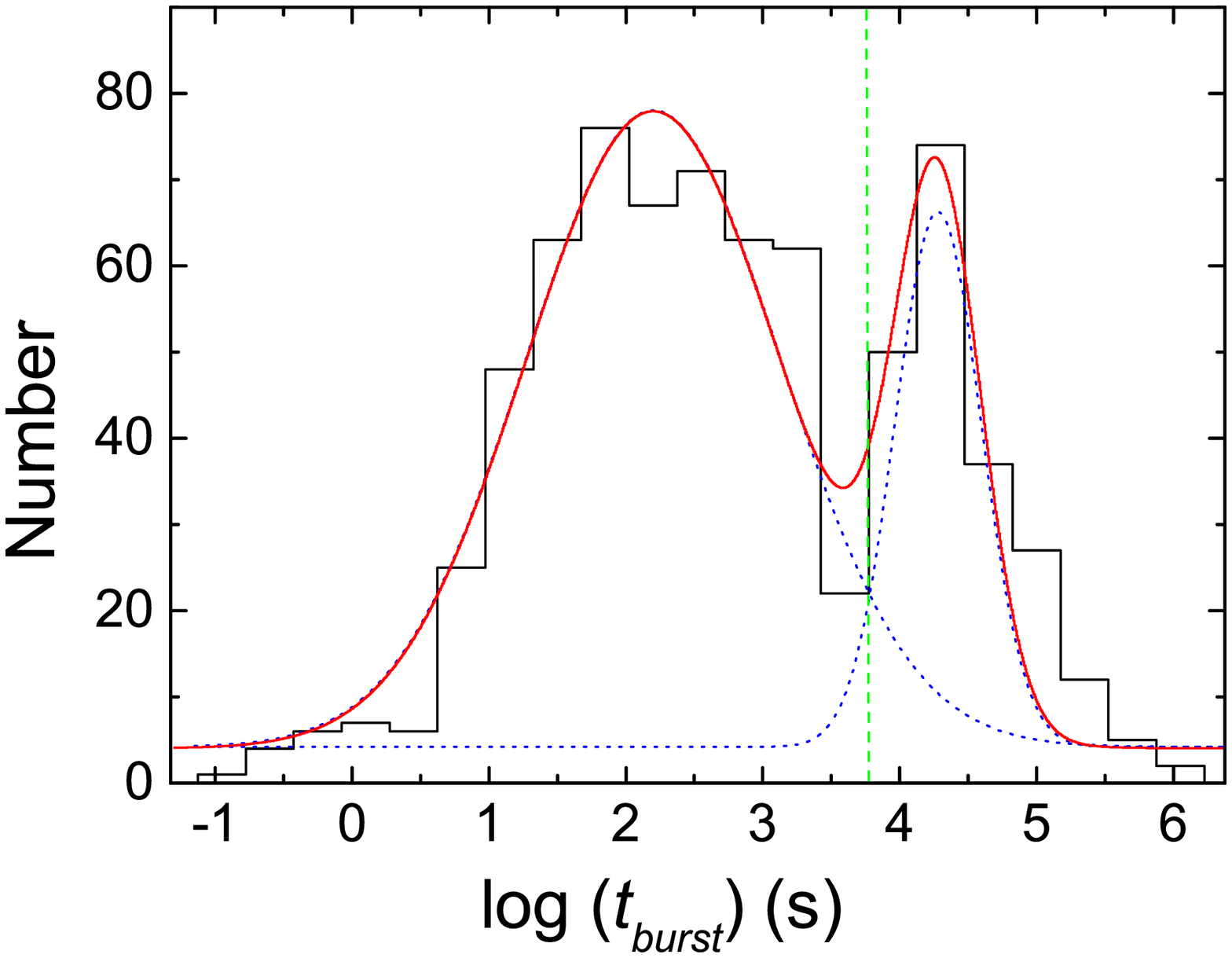}}
       \hspace{-0.4in}
    \subfigure[]{
    \includegraphics[width=2.0in]{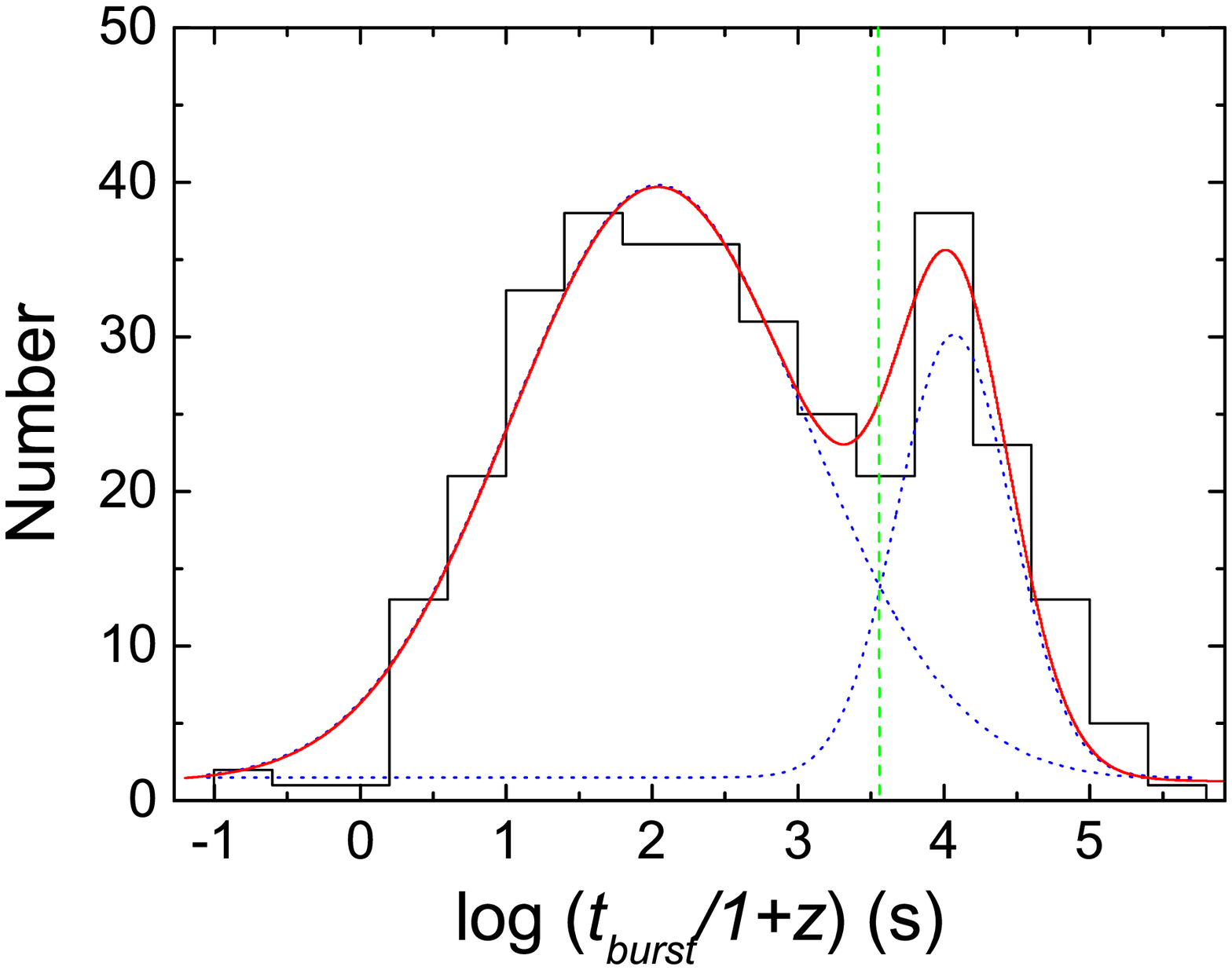}}\\
      \caption{$t_{\rm burst}$ distributions in the observer frame and rest frame for various situations.}
            \end{figure*}

\section{Results}

In \cite{zhang14}, the burst duration $t_{\rm burst}$ is defined as the maximum of $T_{90}$ of $\gamma$-ray emission and the transition time of the last steep-to-shallow transitions in the X-ray light curve. In this case, for our category (1), $t_{\rm burst}$ equals to $T_{90}$. For our category (2), (3) and (4), we have $t_{\rm burst}={\rm max}~(t_{\rm stp}~,~T_{90})$. We plot the distribution of $t_{\rm burst}$ in log space in Fig 1(a). We perform Lilliefors test (i.e., Kolmogorov-Smirnov test for normality with mean and variance unknown, \cite{ lilliefors67,lilliefors69}) on this distribution and it rejects the null hypothesis of normality at the $\leq 0.001$ significance level. We thus fit the distribution with a mixture of two normal distributions in log space, and with chi-square test statistics to find the best-fit parameters. In this case, we fit the distribution with a mixture of two normal distributions, with a narrow, significant peak at $173.8$ s, and a wider, less significant peak at $2.75\times10^{4}$ s, respectively. The division line between the two normal distributions is $t_{\rm burst}=1.86\times10^{4}$ s. In the last column of the Table 1 (Adjusted $R^2$), we show the determinate coefficient of the best fit.

\begin{figure*}
\label{fig:1}
    \subfigure[]{
    \includegraphics[width=2.0in]{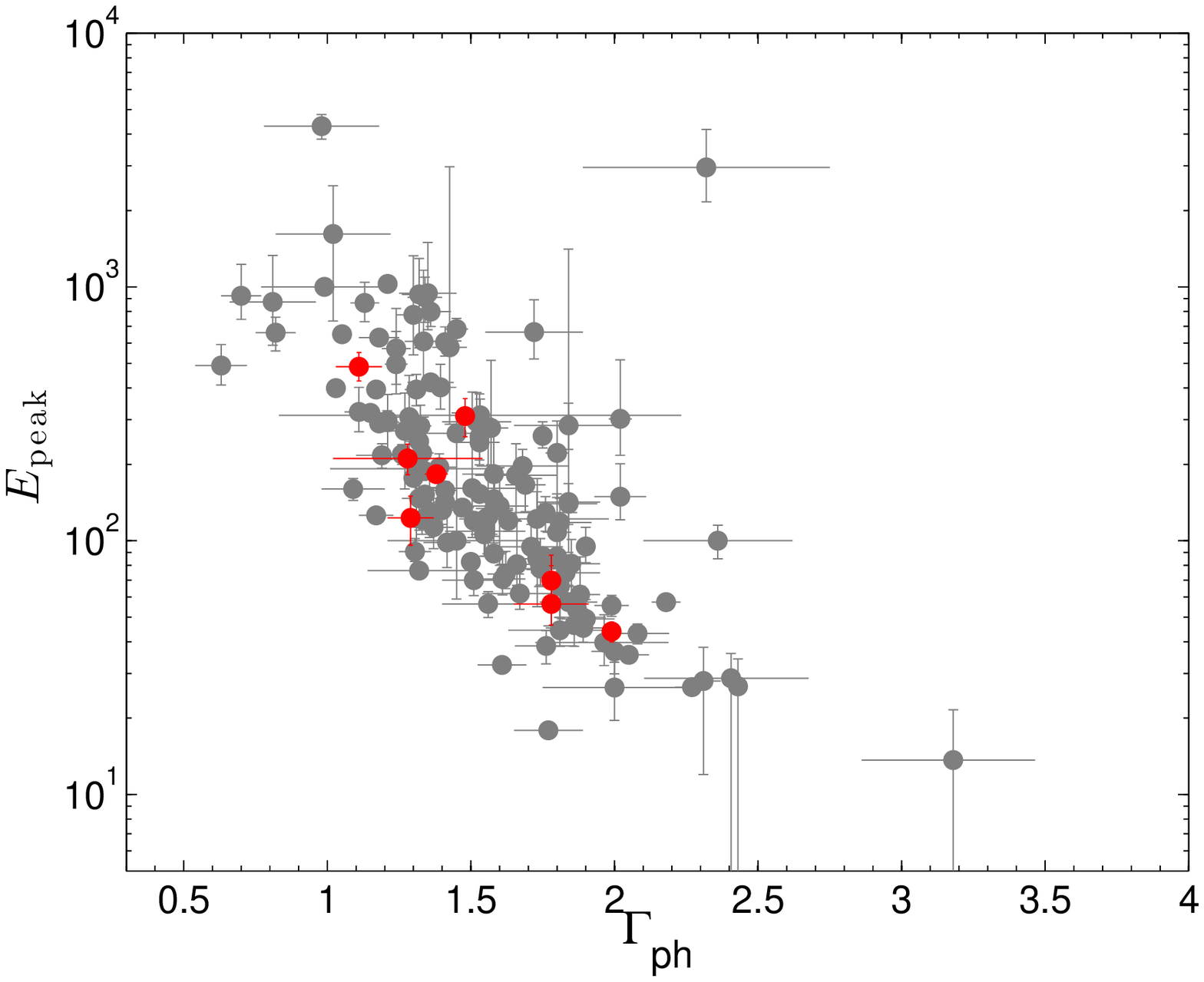}}
    \subfigure[]{
    \includegraphics[width=2.0in]{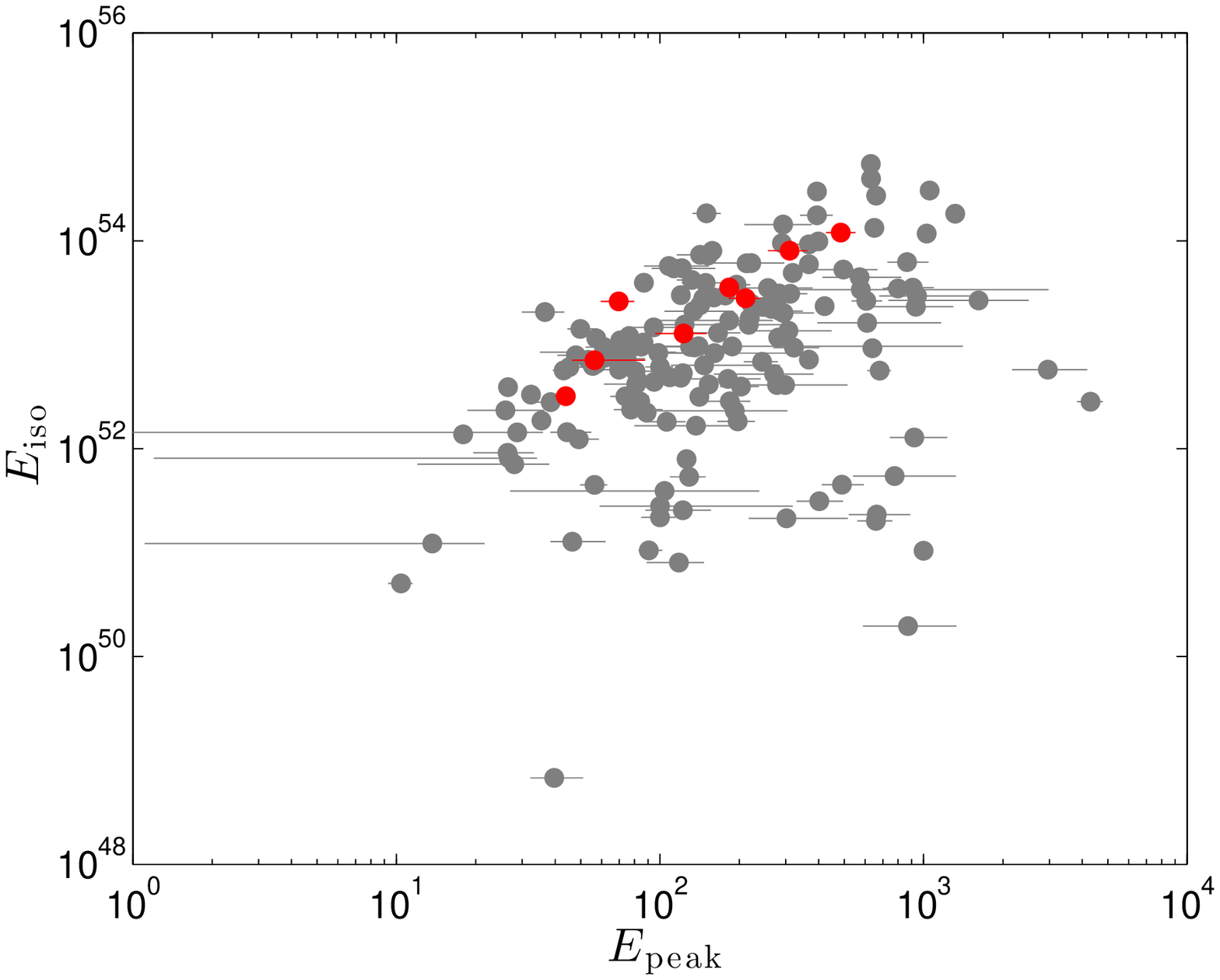}}
     \subfigure[]{
       \includegraphics[width=2.0in]{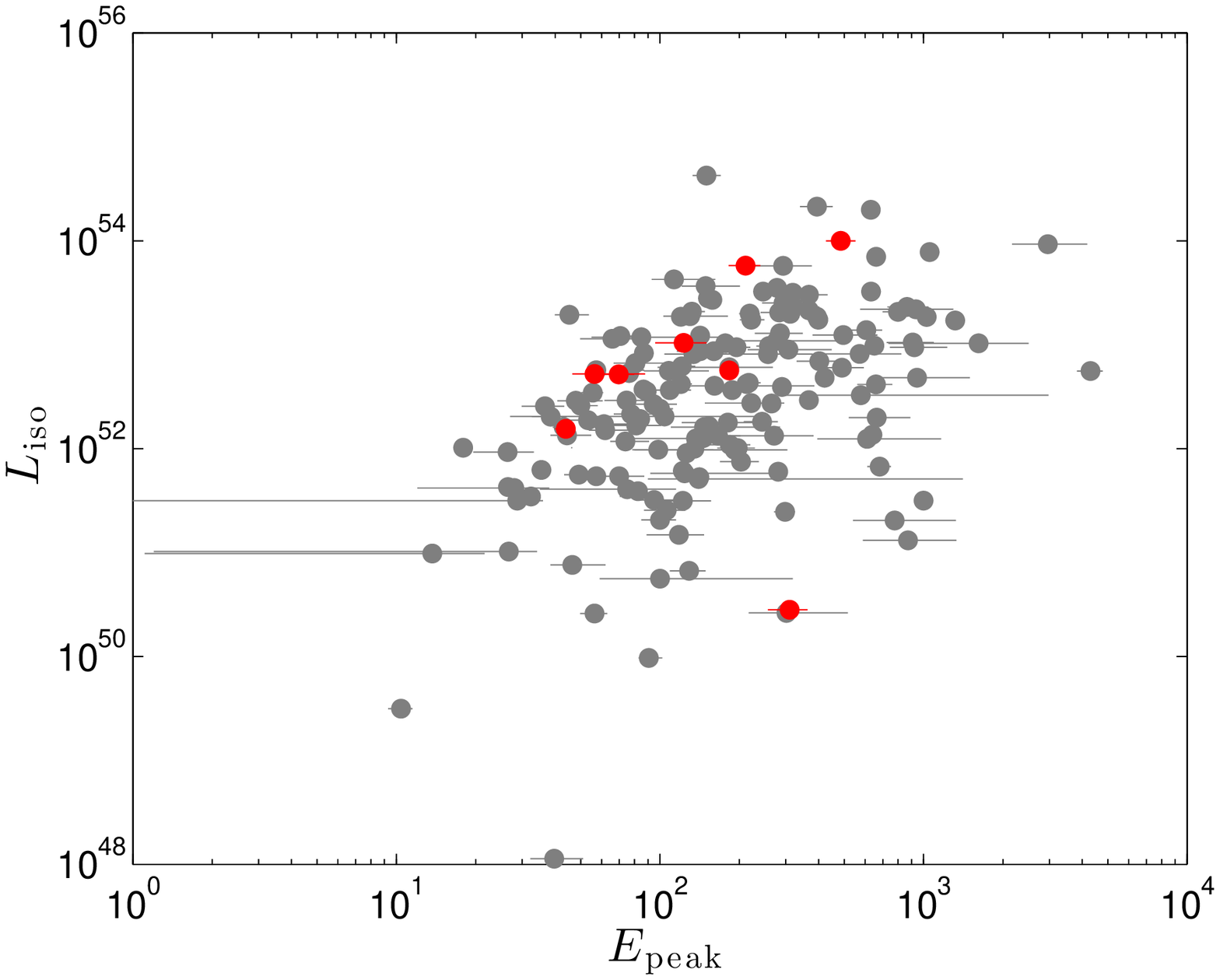}}\\
       \centering
    \subfigure[]{
    \includegraphics[width=2.0in]{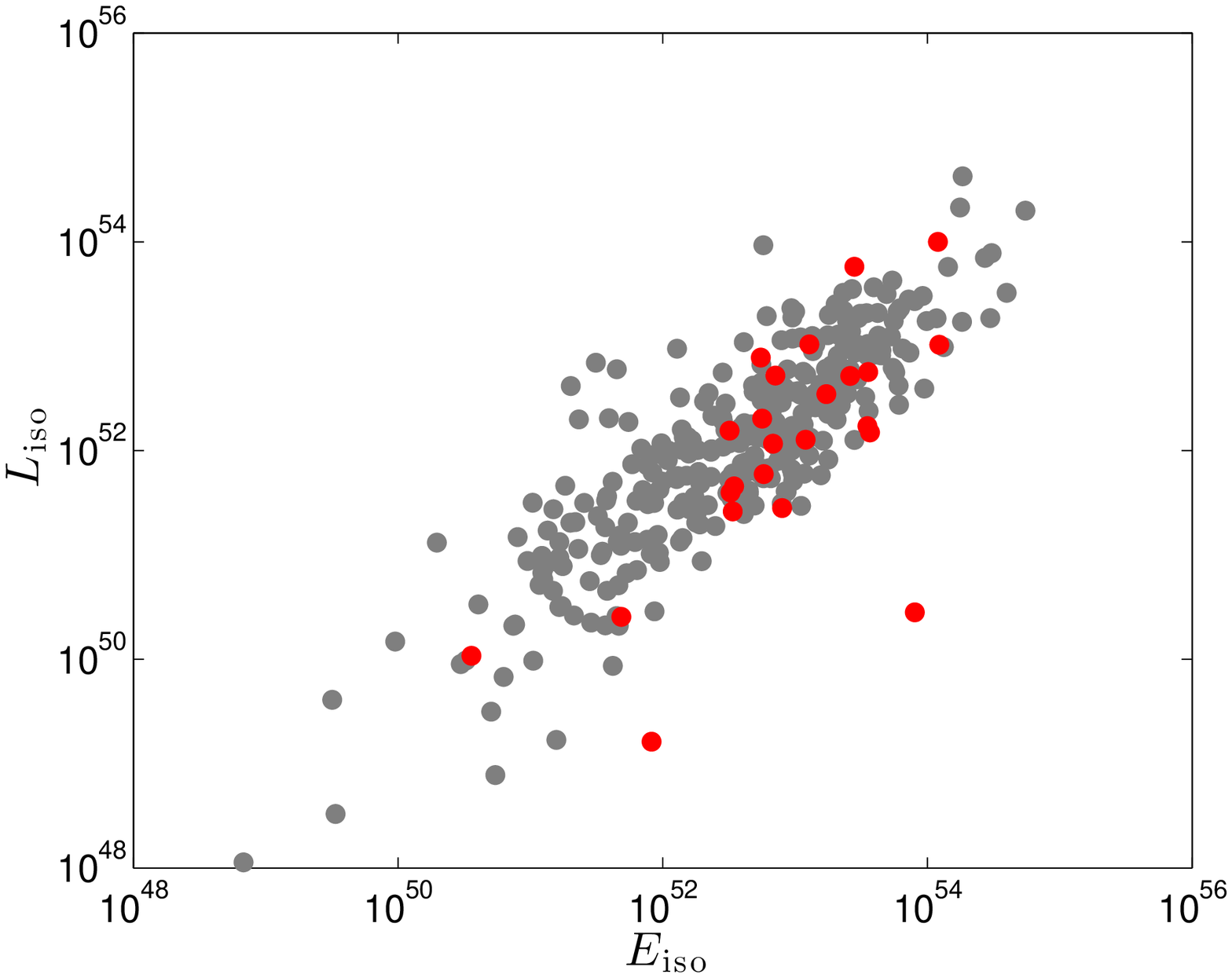}}
     \subfigure[]{
    \includegraphics[width=2.0in]{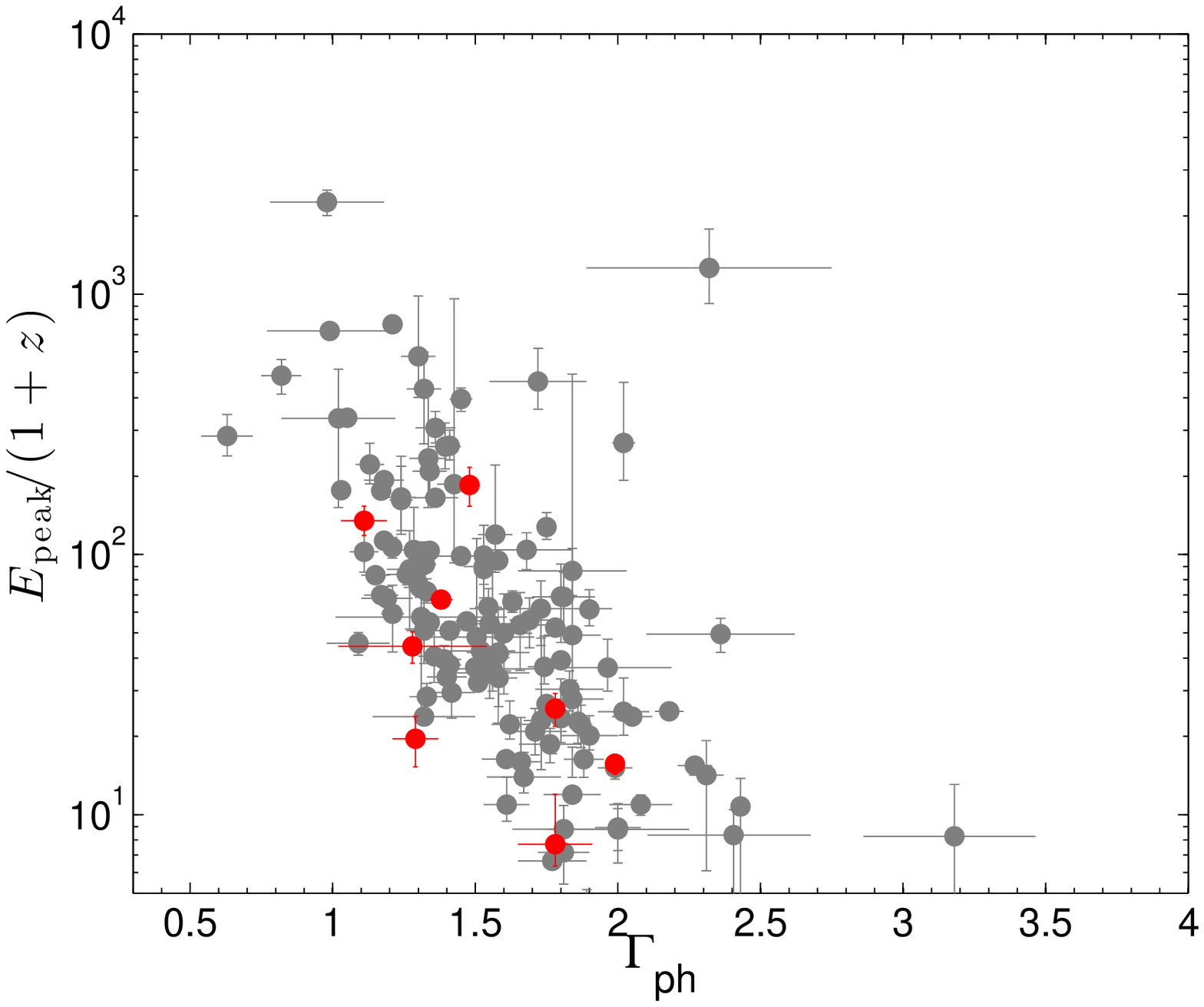}}\\
    \caption{2-D distribution diagrams for prompt emission parameters, such as $E_{p}$, $\Gamma$, $E_{\gamma,\rm iso}$ and $L_{\gamma,\rm iso}$. Red dots present GRBs within the ultra-long population and grey dots present GRBs within the normal population. For all plots, $t_{\rm burst}$ is defined as the maximum of $T_{90}$ of $\gamma$-ray emission and the transition time of the last steep-to-shallow transitions in the X-ray light curve, and the division line between two populations is $t_{\rm burst}=1.86\times10^{4}$ s.}
            \end{figure*}

In our sample, there are 338 GRBs with redshift measurement. For these bursts, the distribution of $t_{\rm burst}$ in rest frame is shown in Fig 1(b). According to Lilliefors test, the distribution rejects the null hypothesis of normality at the $\leq 0.001$ significance level. The distribution can be fitted with a mixture of two normal distributions in log space, with one peaking at $79.4$ s, and the other peaking at $1.66\times10^{4}$ s, respectively. The division line between the two normal distributions is $t_{\rm burst}/(1+z)=6.76\times10^{3}$ s.

\begin{table*}
\begin{center}{\scriptsize
\caption{Fitting results for $t_{\rm burst}$ distributions in the observer frame and rest frame for various situations.}
\begin{tabular}{c|c|c|c|c|c|c|c} \hline\hline
Case & $P_{\rm KS}$ &$\log_{10}{\mu_1}~(s)$ & $\log_{10}{\sigma_1}~(s)$ &$\log_{10}{\mu_2}~(s)$ & $\log_{10}{\sigma_2}~(s)$ &$\log_{10}{T_{\rm div}}~(s)$& Adjusted $R^2$\\
\hline
I (obs) & $\leq0.001$ & $2.24\pm0.06$ &   $1.78\pm0.17$   &  $4.44\pm0.05$   & $0.43\pm0.23$    & $4.27$& 0.92\\
\hline
I (rest) & $\leq0.001$ & $1.90\pm0.06$ &   $1.61\pm0.17$   &  $4.22\pm0.04$   & $0.51\pm0.44$    & $3.83$&0.91\\
\hline
II  (obs)& $\leq0.001$ & $2.20\pm0.05$ &   $1.75\pm0.12$   &  $4.32\pm0.02$   & $0.37\pm0.12$    & $3.99$& 0.96\\
\hline
II  (rest)& $0.004$ & $1.84\pm0.04$ &   $1.69\pm0.12$   &  $3.93\pm0.02$   & $0.45\pm0.09$    & $3.54$& 0.97\\
\hline
III (obs) & $\leq0.001$ & $2.22\pm0.06$ &   $1.79\pm0.16$   &  $4.48\pm0.03$   & $0.66\pm0.26$    & $4.01$& 0.92\\
\hline
III (rest) & $\leq0.001$ & $1.91\pm0.07$ &   $1.74\pm0.20$   &  $4.11\pm0.03$   & $0.66\pm0.33$    & $3.62$& 0.89\\
\hline
IV (obs)& $\leq0.001$ & $2.17\pm0.04$ &   $1.74\pm0.12$   &  $4.31\pm0.03$   & $0.91\pm0.19$    & $3.73$& 0.96\\
\hline
IV (rest)& $0.002$ & $1.81\pm0.05$ &   $1.66\pm0.15$   &  $3.73\pm0.04$   & $0.81\pm0.18$    & $3.22$& 0.94\\
\hline
V (obs)& $0.002$ & $2.36\pm0.08$ &   $1.87\pm0.23$   &  $4.32\pm0.02$   & $0.60\pm0.24$    & $3.97$& 0.90\\
\hline
V (rest)& $\leq0.001$ & $2.09\pm0.06$ &   $1.83\pm0.17$   &  $4.22\pm0.04$   & $0.63\pm0.16$    & $3.77$& 0.95\\
\hline
VI (obs)& $\leq0.001$ & $2.34\pm0.05$ &   $1.97\pm0.17$   &  $4.18\pm0.05$   & $0.99\pm0.01$    & $3.80$& 0.95\\
\hline
VI (rest)& $0.274$ & $2.39\pm0.09$ &   $2.95\pm0.39$   &  $--$   & $--$    & $--$& 0.90\\
\hline
VII (obs)& $\leq0.001$ & $2.28\pm0.06$ &   $1.85\pm0.18$   &  $4.30\pm0.02$   & $0.79\pm0.16$    & $3.83$& 0.93\\
\hline
VII (rest)& $0.009$ & $2.01\pm0.05$ &   $1.74\pm0.14$   &  $3.94\pm0.05$   & $1.23\pm0.22$    & $3.37$& 0.94\\
\hline
VIII (obs)& $\leq0.001$ & $2.20\pm0.05$ &   $1.87\pm0.17$   &  $4.28\pm0.03$   & $0.63\pm0.09$    & $3.74$& 0.93\\
\hline
VIII (rest)& $0.046$ & $2.04\pm0.06$ &   $2.03\pm0.18$   &  $4.07\pm0.03$   & $0.78\pm0.11$    & $3.37$& 0.96\\
   \hline\hline
 \end{tabular}
 }
\end{center}
{\sc Note.} ---
\textbf{Case I}: the definition of $t_{\rm burst}$ is the same with \cite{zhang14}; \textbf{Case III}: the contributions from the late external plateaus  (e.g., $\gtrsim 10^4$ s) are involved when defining $t_{\rm burst}$;  \textbf{Case V}: for all GRBs without X-ray flares but with external plateau, the contributions from the external plateaus are involved when defining $t_{\rm burst}$; \textbf{Case VII}: the contributions from all external plateaus are involved; \textbf{Case II, IV, VI} and \textbf{VIII}: the definition of $t_{\rm burst}$ is the same with Case I, III, V and VII, respectively, but considering the observational gap effect. $\bm{P_{\rm KS}}$ is the P value of Kolmogorov-Smirnov test for normality of the distribution of $t_{\rm burst}$; $\bm{\mu_1}$, $\bm{\sigma_1}$, $\bm{\mu_2}$ and $\bm{\sigma_2}$ are mean and variance values for the best fit of $t_{\rm burst}$ with two normal distributions. $\bm{T_{\rm div}}$ is the division line between the two normal distributions.  $\bm{R^2}$ is the coefficient of determination for the best fit. 
\end{table*}

As discussed in \cite{zhang14}, this apparent bimodal distribution might be subject to strong selection effects due to observational biases, especially the observational gap effect. In general, for XRT GRB observations, there is an observational gap around thousands of seconds due to various reasons such as geometry configuration between the satellite orbital position relative to the GRB source position, instrumental feature of the satellite, and delay of observation in respect to the priority of other ongoing observations. All these factors act as a selection effect against finding $t_{\rm burst}$ values within this gap, which may explain the sudden drop of $t_{\rm burst}$ around $\sim1000$ s. To justify such a selection effect, we systemically go through our entire sample, and we find that for $80\%$ of the bursts, the XRT data before the observational gap could be fitted with a single power law decay, and the power law fitting could be well extrapolated to the observational data after the gap. For $10\%$ of the bursts, the XRT data before the observational gap could be fitted with a single power law decay, while the gap together with the data after the gap could be fitted with another single power law decay. Nevertheless the starting point of the gap is far away from the last flare. For these $90\%$ sources, it should be with small probability to have $t_{\rm burst}$ falling in the observational gap. For the other $10\%$ bursts, there appears X-ray flares before the observational gap, and the ending time of the last flare is very close to the beginning time of the gap. For these sources, it is possible that some later X-ray flares are missed due to the data gap, and the $t_{\rm burst}$ indeed falls into the gap region. For these bursts, we record the ending time of their observational gap as $t_{\rm gap}$. For testing the selection effect, we recalculate the $t_{\rm burst}$ value for these sources as $t_{\rm burst}={\rm max}~(t_{\rm stp}~,~T_{90},~t_{\rm gap})$. We plot the new distribution of $t_{\rm burst}$ and $t_{\rm burst}/(1+z)$ in Fig 1(c) and (d). The Lilliefors test result rejects the null hypothesis of normality of the distribution of $t_{\rm burst}$ at the $\leq 0.001$ significance level and rejects the null hypothesis of normality of the distribution of $t_{\rm burst}/(1+z)$ at the $0.004$ significance level. The distribution of $t_{\rm burst}$ could be fitted with a mixture of two normal distributions in log space, with one peaking at $158.5$ s, and the other peaking at $2.09\times10^{4}$ s, respectively. The division line between the two normal distributions is $t_{\rm burst}=9.77\times10^{3}$ s. The distribution of $t_{\rm burst}/(1+z)$ could be fitted with a mixture of two normal distributions in log space, with one peaking at $69.2$ s, and the other peaking at $8.51\times10^{3}$ s, respectively. The division line between the two normal distributions is $t_{\rm burst}/(1+z)=3.46\times10^{3}$ s. It turns out that the bimodal distribution feature of $t_{\rm burst}$ and $t_{\rm burst}/(1+z)$ is not due to the selection effect. We thus propose that in terms of burst duration $t_{\rm burst}$ or $t_{\rm burst}/(1+z)$ (i.e., central engine activity timescale either in the observer frame and rest frame), a new population of ultra-long GRBs indeed exists.

It is of great interest to investigate whether the bursts within the ultra-long population being statistically different in sense of other features besides the duration distribution. We thus plot our collected prompt emission parameters, such as $E_{p}$, $\Gamma$, $E_{\gamma,\rm iso}$ and $L_{\gamma,\rm iso}$, in pairs in 2-D distribution diagrams (see Fig. 2). We find that except the duration distribution, there is no distinct 2-D distribution plots that can clearly separate the ultra-long population of GRBs from the normal population. 

In case of possible selection biases from \cite{li16}, which compiles all possible data from various sources, we also compare the properties of our ultra long population with a Swift only sample. In the Swift only sample, properties, such as $E_{p}$, $\Gamma$, fluency and peak flux are obtained from Swift GRB table \footnote{\url{$\rm http://swift.gsfc.nasa.gov/archive/grb\_table/$}}, and $E_{\gamma,\rm iso}$ and $L_{\gamma,\rm iso}$ are estimated with Swift-based properties.  We examined the same 2-D plots with these Swift-based properties, and obtained a consistent result with what we show in Figure 2, i.e., ultra long population are not able to be distinguished from these 2-D diagrams. Such conclusion is applicable to all situations discussed in the following. 

According to \cite{gao15}, as long as the ending time is large enough (e.g., $\gtrsim 10^4$ s), external X-ray plateau could also reflect the intrinsic central engine activity. In this work, we suggest to redefine the burst duration $t_{\rm burst}$ by including the contribution from such late external X-ray plateaus. Specifically, for our category (1), $t_{\rm burst}$ equals to $T_{90}$. For our category (2) and (4), we have $t_{\rm burst}={\rm max}~(t_{\rm stp}~,~T_{90})$. Within our category (3), for those 62 GRBs with external plateau and $t_{\rm pla}\gtrsim10^{4}~\rm{s}$, we have $t_{\rm burst}=t_{\rm pla}$. For other 83 GRBs within category (3), we have $t_{\rm burst}={\rm max}~(t_{\rm stp}~,~T_{90})$. In this case, we plot the distribution of $t_{\rm burst}$ in log space in Fig 1(e). According to Lilliefors test, the distribution rejects the null hypothesis of normality at the $\leq 0.001$ significance level. The distribution can be fitted with a mixture of two normal distributions in log space, with one peaking at $166.0$ s, and the other peaking at $3.02\times10^{4}$ s, respectively. The division line between the two normal distributions is $t_{\rm burst}=1.02\times10^{4}$ s. In this case, the distribution of $t_{\rm burst}$ in rest frame is shown in Fig 1(f). According to Lilliefors test, the distribution rejects the null hypothesis of normality at the $\leq 0.001$ significance level. The distribution can be fitted with a mixture of two normal distributions in log space, with one peaking at $81.3$ s, and the other peaking at $1.29\times10^{4}$ s, respectively. The division line between the two normal distributions is $t_{\rm burst}/(1+z)=4.17\times10^{3}$ s. Considering the observational gap effect, the distributions of $t_{\rm burst}$ and $t_{\rm burst}/(1+z)$ are shown in Fig 1(g) and (h). The Lilliefors test result rejects the null hypothesis of normality of the distribution of $t_{\rm burst}$ at the $\leq 0.001$ significance level and rejects the null hypothesis of normality of the distribution of $t_{\rm burst}/(1+z)$ at the $0.002$ significance level. The distribution of $t_{\rm burst}$ could be fitted with a mixture of two normal distributions in log space, with one peaking at $147.9$ s, and the other peaking at $2.04\times10^{4}$ s, respectively. The division line between the two normal distributions is $t_{\rm burst}=5.37\times10^{3}$ s. The distribution of $t_{\rm burst}/(1+z)$ could be fitted with a mixture of two normal distributions in log space, with one peaking at $64.6$ s, and the other peaking at $5.37\times10^{3}$ s, respectively. The division line between the two normal distributions is $t_{\rm burst}/(1+z)=1.66\times10^{3}$ s. 

In principle, even the ending time is not larger than $10^4$ s, the external X-ray plateau could also be due to the late central engine energy injection. If so, within our category (3), for all GRBs with external plateau, we should have $t_{\rm burst}=t_{\rm pla}$. In this case, we plot the distribution of $t_{\rm burst}$ in log space in Fig 1(i). According to Lilliefors test, the distribution rejects the null hypothesis of normality at the $0.002$ significance level. The distribution can be fitted with a mixture of two normal distributions in log space, with one peaking at $229.1$ s, and the other peaking at $2.09\times10^{4}$ s, respectively. The division line between the two normal distributions is $t_{\rm burst}=9.33\times10^{3}$ s. For the sample with redshift measurement, the distribution of $t_{\rm burst}$ in rest frame is shown in Fig 1(j). According to Lilliefors test, the distribution rejects the null hypothesis of normality at the $\leq 0.001$ significance level. The distribution can be fitted with a mixture of two normal distributions in log space, with one peaking at $123.0$ s, and the other peaking at $1.66\times10^{4}$ s, respectively. The division line between the two normal distributions is $t_{\rm burst}/(1+z)=5.89\times10^{3}$ s. Considering the observational gap effect, the distributions of $t_{\rm burst}$ and $t_{\rm burst}/(1+z)$ are shown in Fig 1(k) and (l). The Lilliefors test result rejects the null hypothesis of normality of the distribution of $t_{\rm burst}$ at the $\leq 0.001$ significance level. The distribution of $t_{\rm burst}$ could be fitted with a mixture of two normal distributions in log space, with one peaking at $218.8$ s, and the other peaking at $1.51\times10^{4}$ s, respectively. The division line between the two normal distributions is $t_{\rm burst}=6.31\times10^{3}$ s. 

It is worth noting that in this case, the distribution of $t_{\rm burst}/(1+z)$ could accept the null hypothesis of normality at the $0.274$ significance level. The distribution of $t_{\rm burst}/(1+z)$ could be fitted with a single normal distribution in log space, peaking at $245.5$ s. Such result implies that if all external X-ray plateau is generated by the late central engine energy injection, and if the $t_{\rm burst}$ indeed falls into the observational gap region for some GRBs (e.g., GRBs containing X-ray flares before the observational gap and the ending time of the last flare is very close to the beginning time of the gap), the bimodal distribution of $t_{\rm burst}$ doesn't exist, namely ultra-long GRBs are the tail of the distribution of normal LGRBs rather than corresponding to a new possible population, which would be consistent with the fact that no distinct 2-D distribution plots could clearly separate the ultra-long population of GRBs from the normal population. 

In our category (4), 50 bursts also contain external plateau at late time. However, the external plateaus in these cases 
could be naturally explained without invoking late central engine activity,  since for GRBs with X-ray flares, it is believed that the late ejecta that causes the flares would continue proceeding to overtake and refresh the afterglow shock, thus causing additional activity at even later times in the light curve. Although the possibility is low, for completeness, we also test the situation that these external plateau reflecting central engine activity. In this case, for category (4), we have 
$t_{\rm burst}={\rm max}~(t_{\rm stp}~,~T_{90}~,~t_{\rm pla})$. We plot the new distribution of $t_{\rm burst}$ in log space in Fig 1(m). According to Lilliefors test, the distribution rejects the null hypothesis of normality at the $\leq 0.001$ significance level. The distribution can be fitted with a mixture of two normal distributions in log space, with one peaking at $190.5$ s, and the other peaking at $2.00\times10^{4}$ s, respectively. The division line between the two normal distributions is $t_{\rm burst}=6.76\times10^{3}$ s. For the sample with redshift measurement, the distribution of $t_{\rm burst}$ in rest frame is shown in Fig 1(n). According to Lilliefors test, the distribution rejects the null hypothesis of normality at the $0.009$ significance level. The distribution can be fitted with a mixture of two normal distributions in log space, with one peaking at $102.3$ s, and the other peaking at $8.71\times10^{3}$ s, respectively. The division line between the two normal distributions is $t_{\rm burst}/(1+z)=2.34\times10^{3}$ s.  Considering the observational gap effect, the distributions of $t_{\rm burst}$ and $t_{\rm burst}/(1+z)$ are shown in Fig 1(o) and (p). The Lilliefors test result rejects the null hypothesis of normality of the distribution of $t_{\rm burst}$ at the $\leq 0.001$ significance level and rejects the null hypothesis of normality of the distribution of $t_{\rm burst}/(1+z)$ at the $0.046$ significance level. The distribution of $t_{\rm burst}$ could be fitted with a mixture of two normal distributions in log space, with one peaking at $158.5$ s, and the other peaking at $1.91\times10^{4}$ s, respectively. The division line between the two normal distributions is $t_{\rm burst}=5.50\times10^{3}$ s. The distribution of $t_{\rm burst}/(1+z)$ could be fitted with a mixture of two normal distributions in log space, with one peaking at $109.6$ s, and the other peaking at $1.17\times10^{4}$ s, respectively. The division line between the two normal distributions is $t_{\rm burst}/(1+z)=2.34\times10^{3}$ s. Note that for those 50 GRBs containing external plateau, 36 of them are with $t_{\rm pla} \gtrsim 10^4$ s and 14 are with $t_{\rm pla} < 10^4$ s. For those 14 bursts, most of their $t_{\rm pla}$ values are only slightly smaller than $10^4$ s, so that either or not separating out these 14 bursts did not make too much difference for the results, that is why we did not present the results by separating these 50 bursts as what we did in group 3.

\section{Conclusion and discussion}

It has been widely discussed that the true GRB central engine activity duration, $t_{\rm burst}$, should be defined by considering both $\gamma$-ray and X-ray data. In principle, the prompt $\gamma$-ray emission, X-ray flares, internal X-ray plateau and even external X-ray plateau could be all originated from internal dissipation, which essentially reflect the intrinsic central engine activity. In this work, we systematically investigate 1032 Swift GRBs that were detected by XRT from January 2005 to June 2016, and we did the following investigations:

Following \cite{zhang14}, the definition of  $t_{\rm burst}$ would be the maximum of $T_{90}$ of $\gamma$-ray emission and the transition time of the last steep-to-shallow transitions in the light curve, which essentially incorporates the prompt $\gamma$-ray emission, X-ray flares and internal X-ray plateau. In this case, we find that both in the observer frame and in the rest frame, the bimodal distribution feature of $t_{\rm burst}$ and $t_{\rm burst}/(1+z)$ indeed exists and the bimodal feature is not caused by the observational biases, e.g., the observational gap effect. To investigate whether the bursts within the ultra-long population being statistically different in sense of other features besides the duration distribution, we plot the prompt emission parameters, such as $E_{p}$, $\Gamma$, $E_{\gamma,\rm iso}$ and $L_{\gamma,\rm iso}$, in pairs in 2-D distribution diagrams and we find that except the duration distribution, there is no distinct 2-D distribution plots that can clearly separate the ultra-long population of GRBs from the normal population. 

On the other hand, we suggest to redefine the burst duration $t_{\rm burst}$ by including the contributions from the external X-ray plateaus. Among our sample, there are 135 bursts consisting external plateaus but without X-ray flares, 47 bursts showing X-ray flares and additional external plateau feature after all flares. For the external plateau sample, we have 62 GRBs with $t_{\rm pla}\gtrsim10^{4}~\rm{s}$, and 73 GRBs with $t_{\rm pla}<10^{4}~\rm{s}$. According to \cite{gao15}, as long as the ending time is large enough (e.g., $\gtrsim 10^4$ s), external X-ray plateau could surely reflect the intrinsic central engine activity. However, for external plateaus whose ending time is relatively small (e.g., $< 10^4$ s), the plateau phase might be due to the late central engine energy injection, but it also has chance to be due to the internal collisions or refreshed external collisions from early ejected shells, whereas the external plateau phase no longer reflects the intrinsic central engine activity. For those 47 bursts showing X-ray flares and additional external plateaus, the possibility is even lower that these external plateau reflecting central engine activity, since for GRBs with X-ray flares, it is believed that the late ejecta that causes the flares would continue proceeding to overtake and refresh the afterglow shock, thus causing additional activity at even later times in the light curve.

In this work, we first only involve the contributions from the late external plateaus  (e.g., $\gtrsim 10^4$ s), we find that the bimodal distribution feature of $t_{\rm burst}$ and $t_{\rm burst}/(1+z)$ becomes more significant, and the bimodal feature would not disappear when the observational gap effect is considered. Secondly, for all GRBs without X-ray flares but with external plateau, we involve the contributions from the external plateaus, and we find that the bimodal distribution feature of $t_{\rm burst}$ and $t_{\rm burst}/(1+z)$ still exists, but the distribution of $t_{\rm burst}/(1+z)$ could accept the null hypothesis of normality (i.e., could be fitted with a single normal distribution in log space) at the $0.274$ significance level when the observational gap effect is considered. Finally, for GRBs with both X-ray flares and external plateau, we also involve the contributions from the external plateaus, and we find that the bimodal distribution feature of $t_{\rm burst}$ and $t_{\rm burst}/(1+z)$ always exists, even when the observational gap effect is considered.

Based on the results of our investigations, the following physical implications could be inferred:

For all situations, the distribution of $t_{\rm burst}$ of GRBs requires two normal distributions in log space to 
provide a good fit, both in observer frame and rest frame. Considering the observational gap effect would not completely erase the bimodal distribution feature. The bimodal feature may suggest that an ultra-long population indeed exist, at least in regard to duration term. However, no distinct 2-D distribution plots of prompt parameters could clearly separate the ultra-long population of GRBs from the normal population meaning that the bursts within the ultra-long population may have no statistically different in sense of other features besides the duration term. To reconcile these two results, we suggest that if the ultra-long population of GRBs indeed exists, their central engine mechanism and radiation mechanism should be similar to the normal population, but they somehow have longer central engine activity timescale. Under the framework of the collapsar model, the central engine (black hole) activity timescale could have a wide range, depending not only on the size of the progenitor star but also on the stellar structure and rotation rate of the progenitor star \citep{kumar08a,kumar08b}. Invoking a larger size of progenitor star, such as a blue supergiant-like progenitor for ultra-long GRBs could naturally explain the unusually long duration. However, \cite{greiner15} recently reported the first discovered association between an ultra-long GRB and a supernova, i.e., GRB 111209A/SN 2011kl. Based on the observed properties of SN 2011kl, such as its spectra and light curve shape, they rule out a blue supergiant progenitor interpretation for GRB 111209A. 

Alternatively, a collapsar model with fallback accretion has been proposed to interpret the ultra-long population of GRBs, and have been successfully applied to fit the broadband data of some typical ultra-long GRBs, such as GRB 121027A \citep{wu13} and GRB 111209A/SN 2011kl \citep{gao16}. In general, the size of the progenitor star for the two population might be similar, but the stellar structure and rotation rate of the progenitor star may be different. The stellar structure and rotation rate could affect the fallback process of the envelope material, which could largely extend the central engine activity time, having a chance to give rise to the ultra-long GRBs. On the other hand, the bounding shock responsible for the associated SN and the baryon-rich wide wind/outflow through the Blandford-Payne (Blandford \& Payne 1982) mechanism from the initial accretion disk would transfer kinetic energy to the envelope materials. If the injected kinetic energy is less than the potential energy of the envelop material, the starting time of the fallback would be delayed, which may even prolong the burst duration. However, if the injected kinetic energy is larger, which might be the majority of cases, the fallback process is vanished and the central engine activity is relatively short, corresponding to the normal long GRBs \citep{gao16}.

Note that for one situation, namely that assuming all external X-ray plateau is generated by the late central engine energy injection (but not involving the external plateau after flares), and if the $t_{\rm burst}$ indeed falls into the observational gap region for some GRBs (e.g., GRBs containing X-ray flares before the observational gap and the ending time of the last flare is very close to the beginning time of the gap), the bimodal feature of $t_{\rm burst}$ could be erased by the observational gap effect. It is still possible that ultra-long GRBs are the tail of the distribution of normal LGRBs rather than corresponding to a new possible population, which would be consistent with the fact that no distinct 2-D distribution plots could clearly separate the ultra-long population of GRBs from the normal population. 

Finally, it is worth noticing that besides observational gap effect, there are some other selection effects might affect the determination of $t_{\rm burst}$, such as the sensitivity of XRT, and the observation ceasing in respect to the priority of other ongoing observations (other GRBs or target of opportunities). These effects could cause underestimating of $t_{\rm burst}$.  Since the XRT observation ending time for $>90\%$ GRBs in our sample is larger than $10^{4}$ s, the underestimating of $t_{\rm burst}$ could only make the ultra-long population even significant and the bimodal distribution of $t_{\rm burst}$ more clear.

\acknowledgments
We thank the referee for the helpful comments which have helped us to improve the presentation
of the paper. This work is supported by the National Basic Research Program ('973' Program) of China (grants 2014CB845800), the National Natural Science Foundation of China under grants 11543005, 11603003, 11603006,11633001, U1431124, 11361140349 (China-Israel jointed program). H.G. acknowledges supported by the Strategic Priority Research Program of the Chinese Academy of Sciences, Grant No. XDB23040100L. H. J. acknowledges support by the Guangxi Science Foundation (grant No. 2016GXNSFCB380005） and One-Hundred-Talents Program of Guangxi colleges. BBZ acknowledges support from the Spanish Ministry Projects AYA 2012-39727-C03-01 and AYA2015-71718-R. The computation resources used in this work are owned by Scientist Support LLC. This work made use of data supplied by the UK Swift Science Data Centre at the University of Leicester. 

\appendix

Multivariate adaptive regression splines (MARS) technique \citep{friedman91} is a non-parametric regression technique that could automatically determine both variable selection and functional form, resulting in an explanatory predictive model. Such MARS model is a linear combination of piecewise polynomial basis functions and can be expressed as 

\begin{eqnarray}
{\hat  {f}}(x)=\sum _{{i=1}}^{{k}}c_{i}B_{i}(x),
\end{eqnarray}
where $B_{i}$ is basis functions, $c_i$  is a constant coefficient. For the 1-order MARS model (in our application) ,  $B_i$  takes one of the following there forms:

\begin{itemize}
\item a constant 1 
\item a Hinge function $\max(0,x-c)$
\item a Hinge function $\max(0,c-x)$
\end{itemize}

\begin{figure}
\centering
    \includegraphics[width=3.0in]{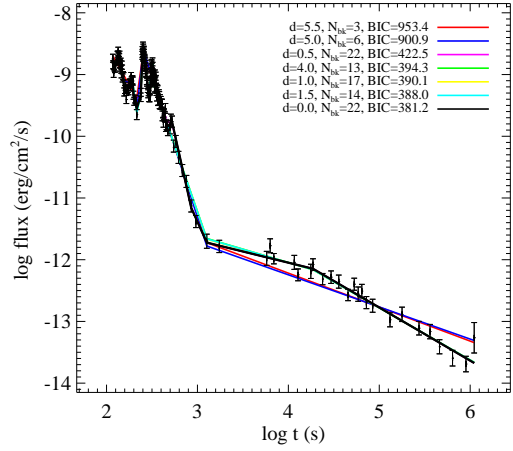}
    \caption{Example (GRB 100302A) of model selection in XRT light curve fitting with MARS technique.} 
    \label{fig:MARS}
            \end{figure}

Those segments are joined together at the knots (which are breaks in the light curve in our case) . To fit data with the MARS model, the procedure first try to repeatedly add basis function to give the maximum reduction in sum-of-squares residual error (so called $\chi^2$ fitting method), then prunes the model by deleting the least effective term at each step until it finds the best sub-model. The latter step involves with a ``penalty" parameter, d, which controls the number of segments and smoothness of the MARS model. In this work, we tried 20 values of d (uniformed distributed from 0 to 10). We compare the outcome models and chose the best one by using the Bayesian information criterion (BIC)  method \cite[][for a detailed overview of MARS technique]{Kweku-Muata Osei-Bryson}. An example of showing how d affects the fitting and how we chose the best model is presented in Figure \ref{fig:MARS}. 

\begin{figure*}
\centering
\label{fig:MARSvsonline}
    \subfigure[]{
    \includegraphics[width=3.0in]{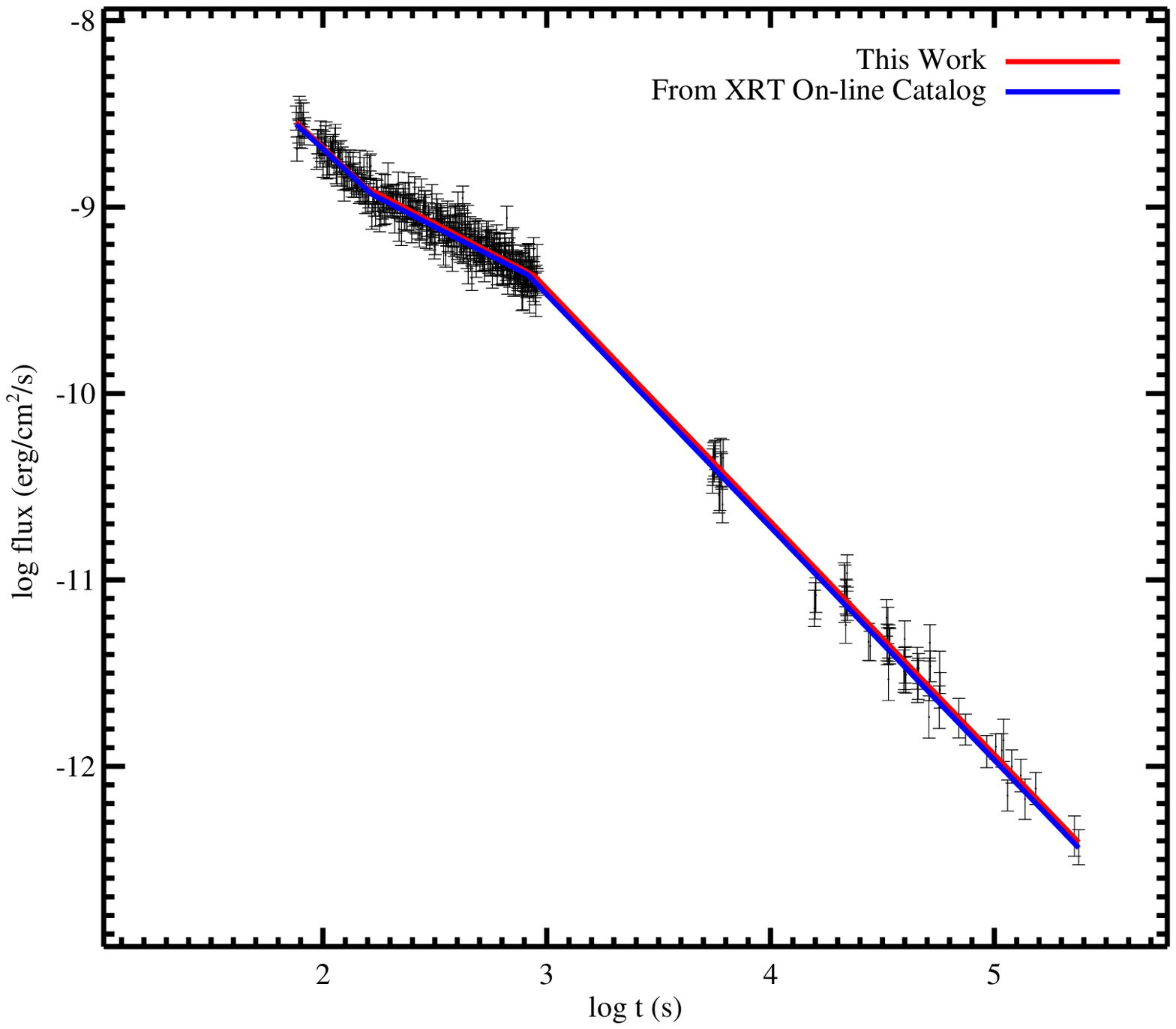}}
    \subfigure[]{
    \includegraphics[width=3.0in]{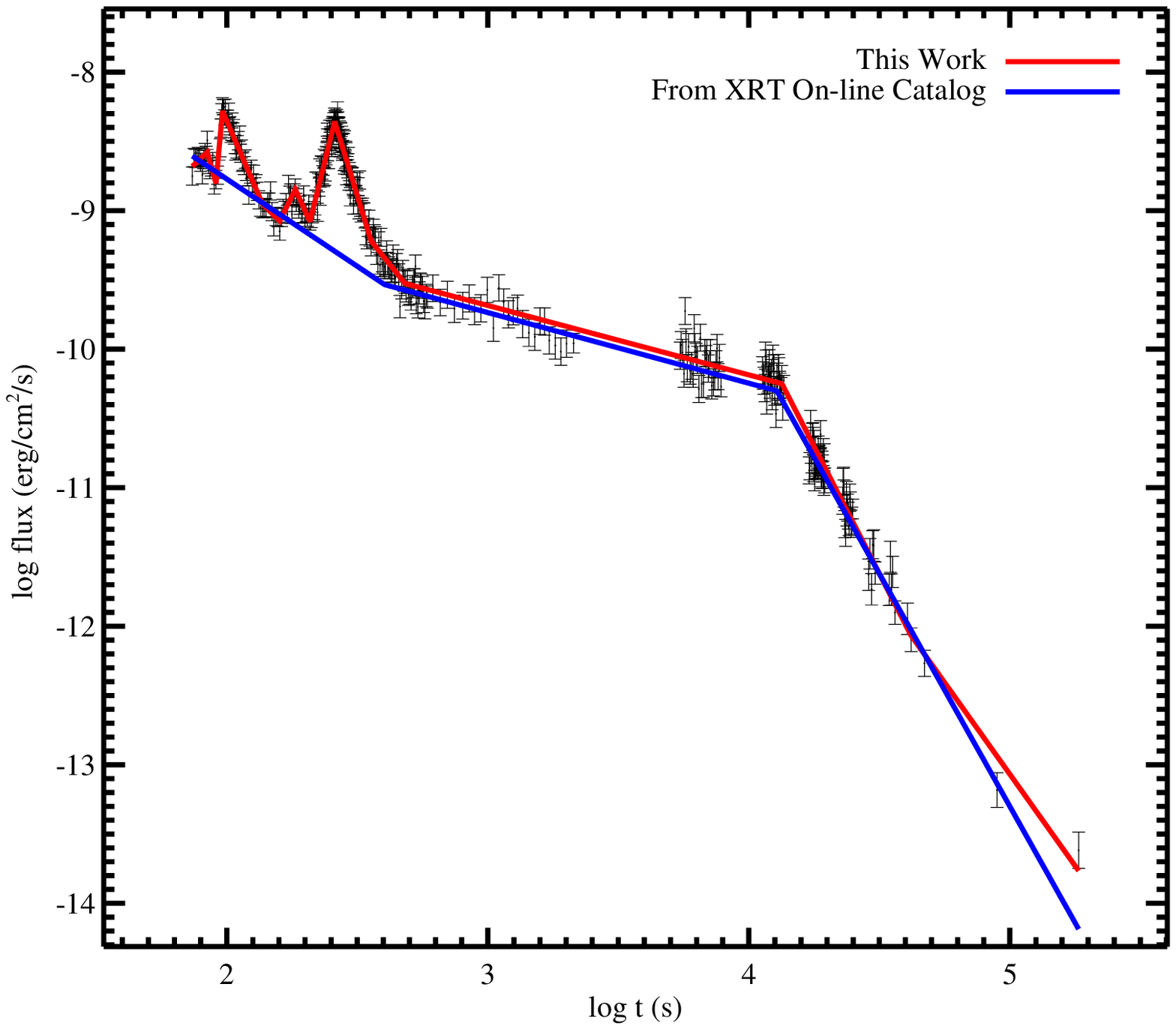}}
    \caption{Comparison of light curve fitting results between MARS technique and the XRT GRB online catalog \citep{evans09}. The left panel is for GRB 150201A, and the right panel is for GRB 060607A. For both panels, the blue line presents the results from XRT online catalog and the red line presents the results from MARS technique.} 
            \end{figure*}

In Figure \ref{fig:MARSvsonline}, we plot two examples of light curve fitting results from MARS technique and from the XRT GRB online catalog \citep{evans09}. It is shown that for bursts without flares (e.g. 150201), the results from MARS technique are well matched with the results from XRT GRB online catalog. For bursts with flares (e.g. 060607), the XRT online catalog only fits the light curve by removing the flare features, while MARS technique could also incorporate the flare phases. It is worth noticing that for these bursts, without considering the flare phase, the fitting result from MARS technique are still consistent with the XRT online catalog. For all bursts in our sample, we list the model information (such as fitting parameters include breaks and slopes) in our on-line real-time HTML table at \url{$\rm http://astrowww.bnu.edu.cn/NewCN/grb/GRB\_XRAY\_FIT/$}. We've made a link for each burst to the UK website for the comparison of our results with XRT online catalog.

In Figure 5, we show some examples of our fitting results for different types of light curves, with marks of $t_{\rm stp}$ and $t_{\rm pla}$ (if applicable), to better illustrate the light curve properties for each of the four categories.

\begin{figure*}
\centering
    \subfigure[]{
    \includegraphics[width=2.0in]{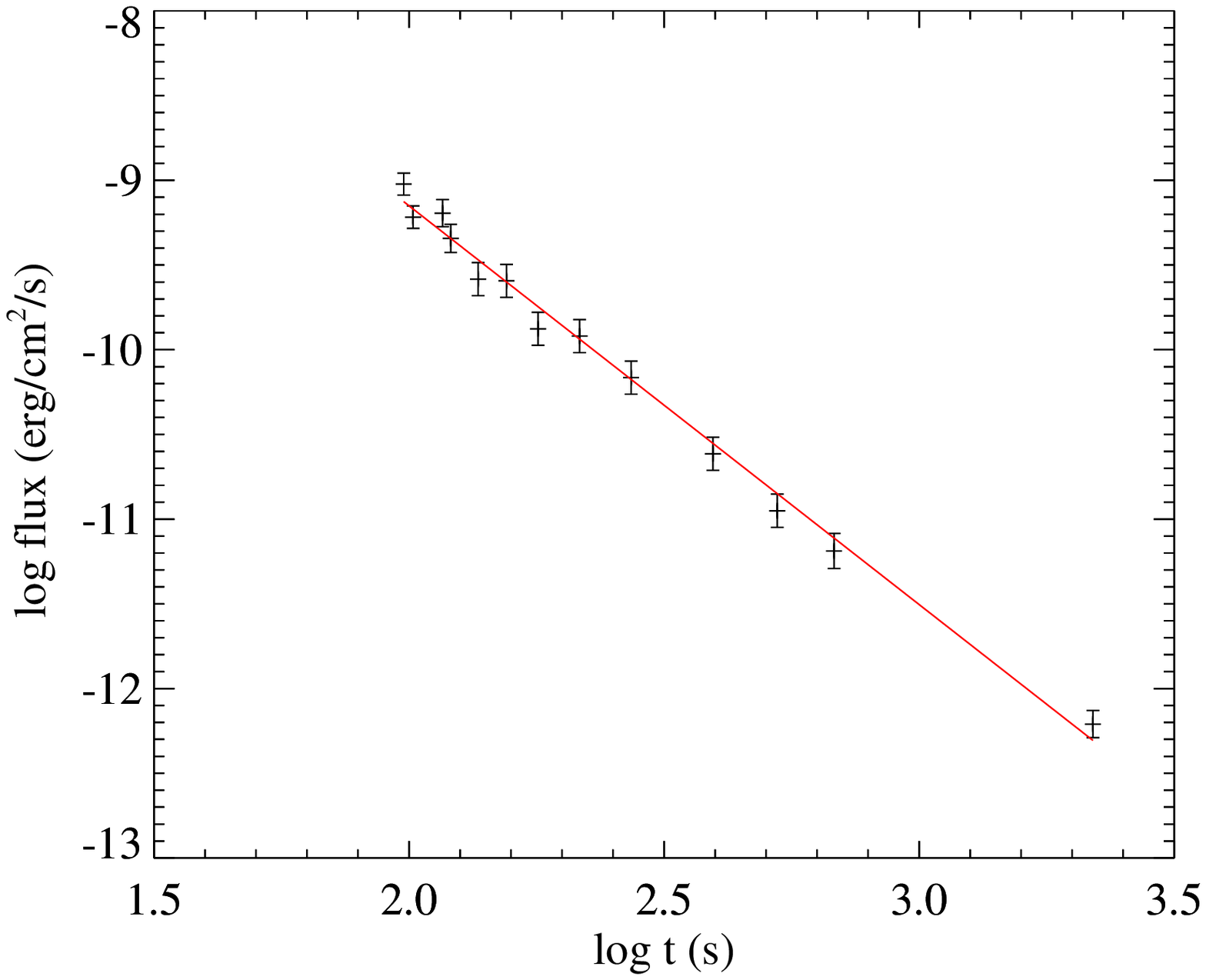}}
    \subfigure[]{
    \includegraphics[width=2.0in]{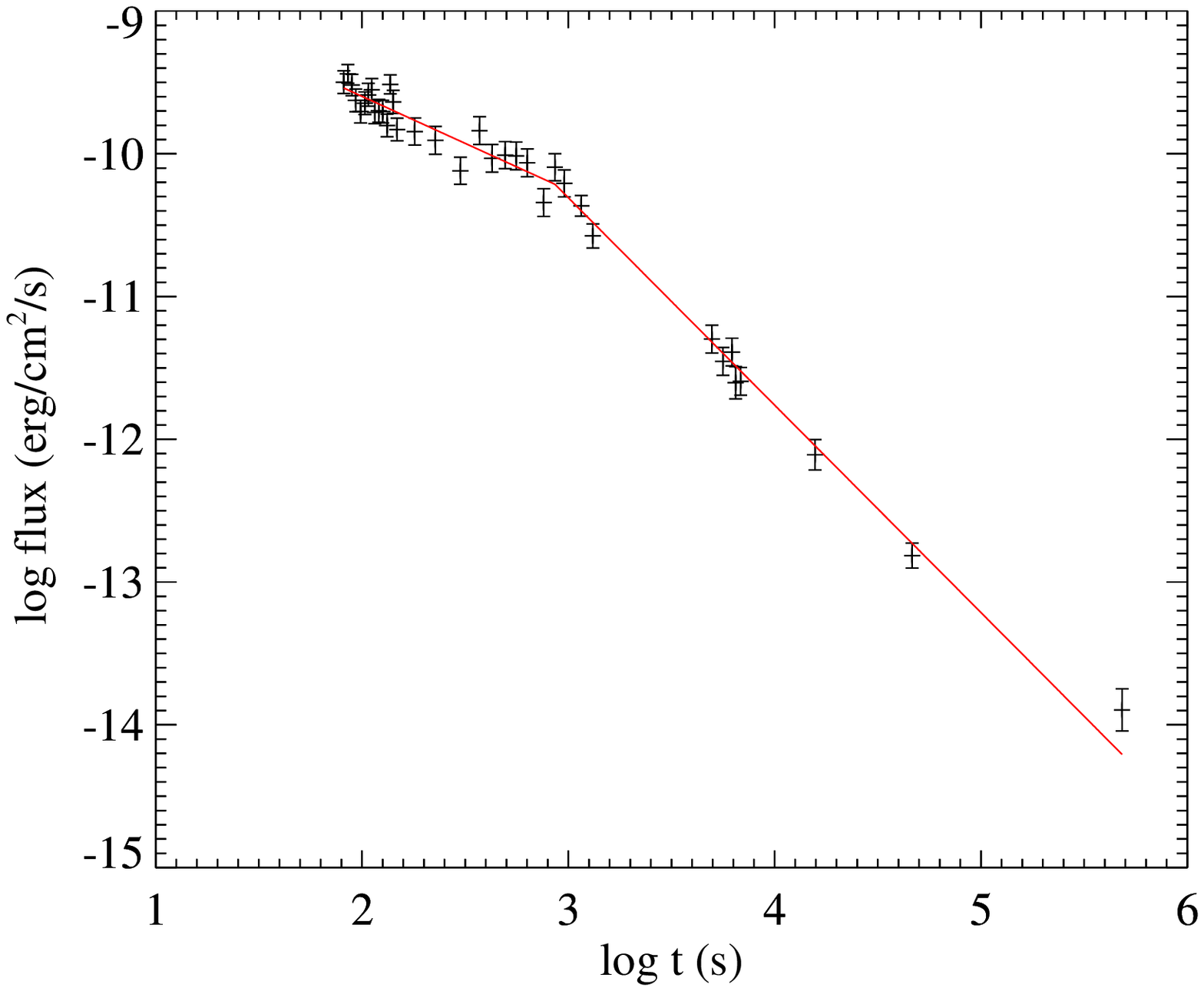}}\\
    
     \subfigure[]{
       \includegraphics[width=2.0in]{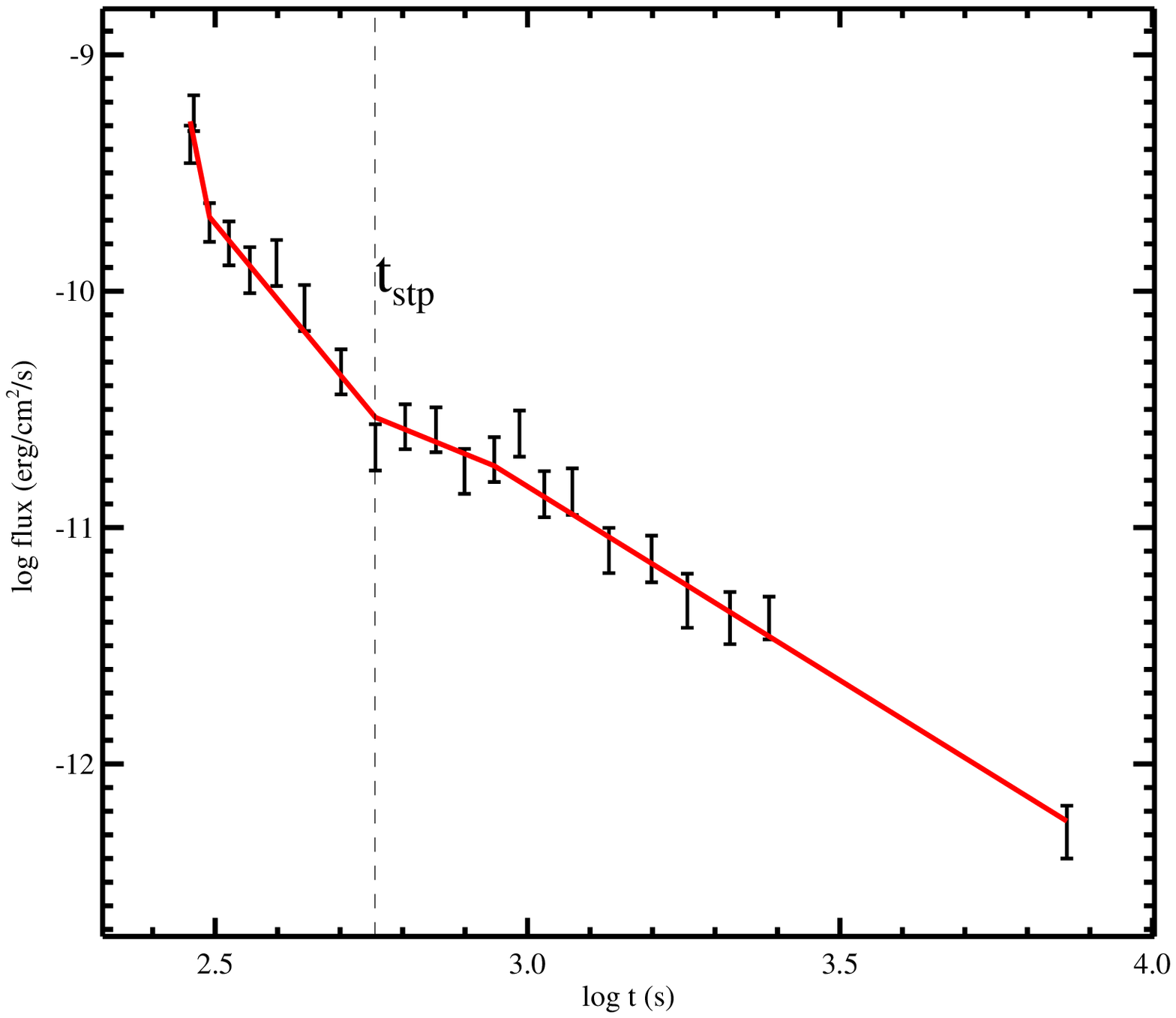}}
    \subfigure[]{
    \includegraphics[width=2.0in]{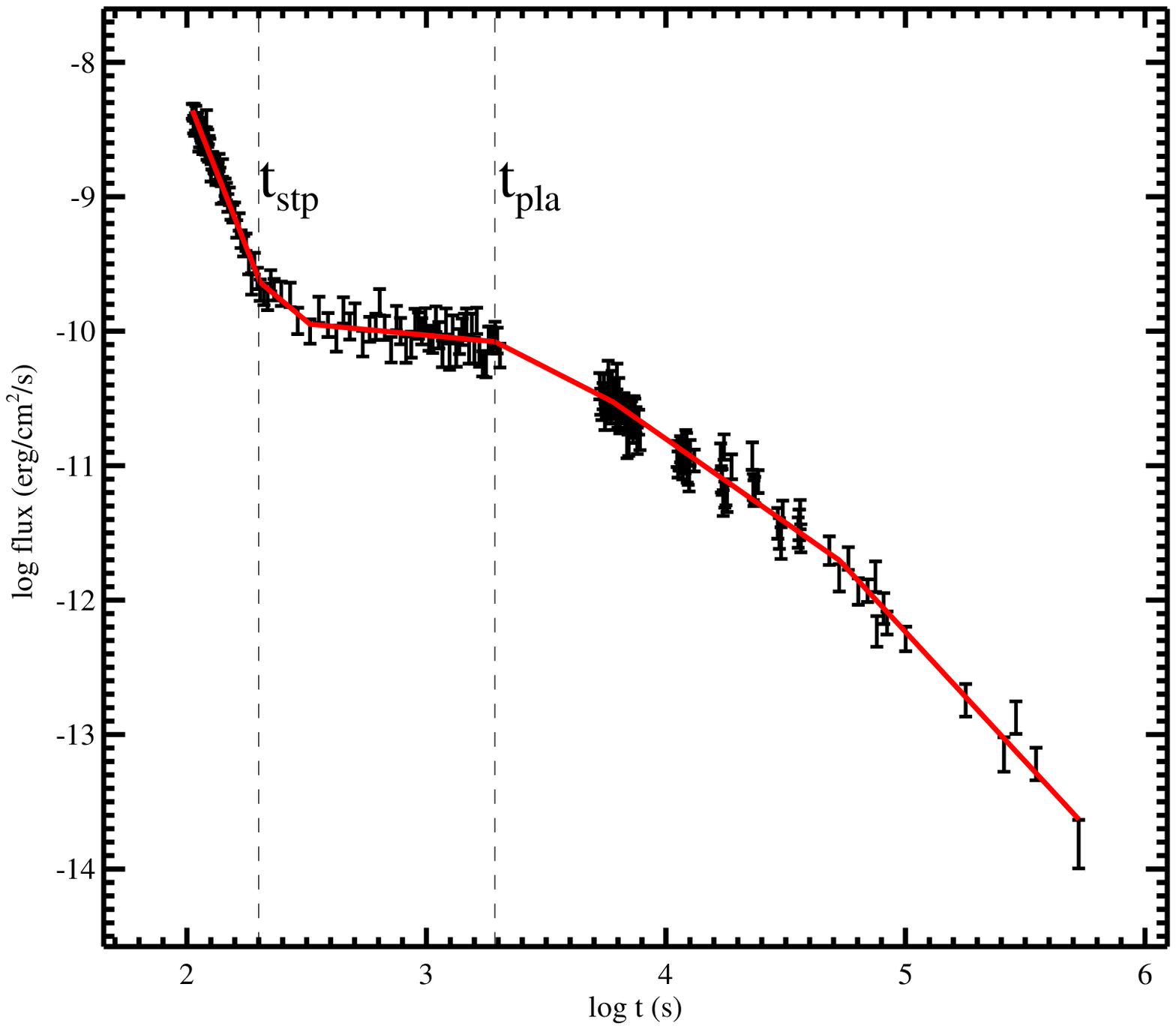}}
    \subfigure[]{
    \includegraphics[width=2.0in]{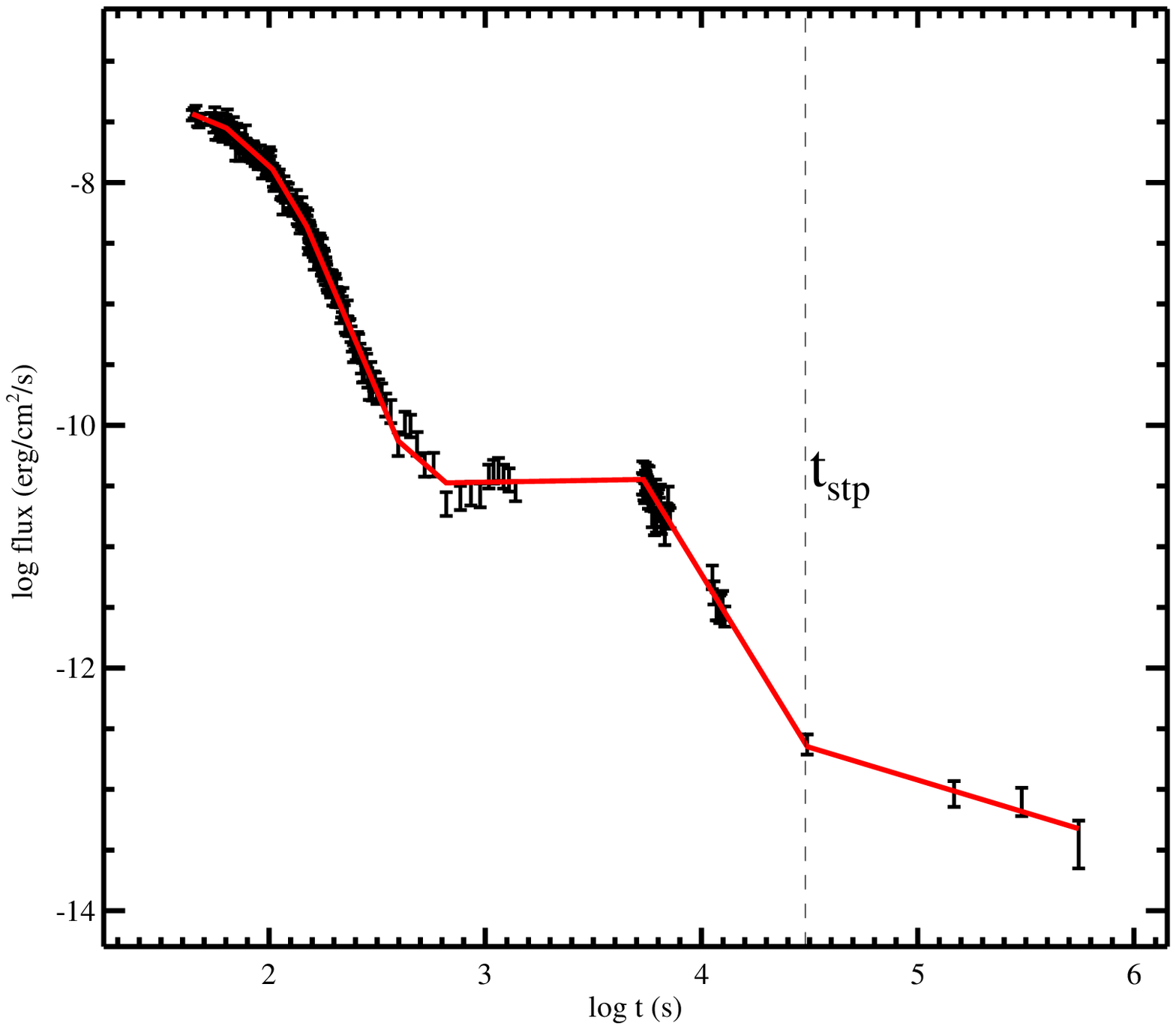}}\\
    \subfigure[]{
    \includegraphics[width=2.0in]{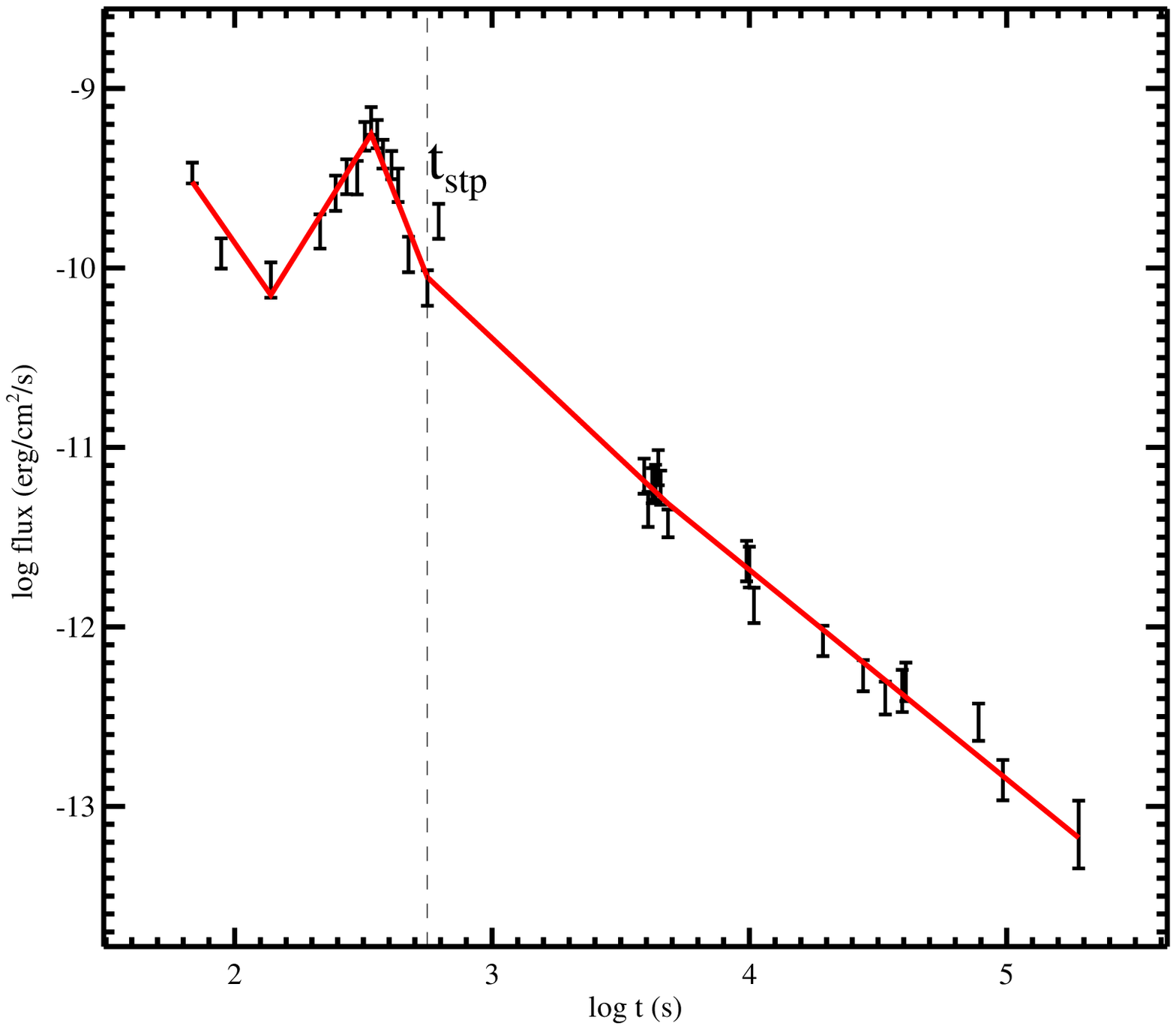}}
    \subfigure[]{
    \includegraphics[width=2.0in]{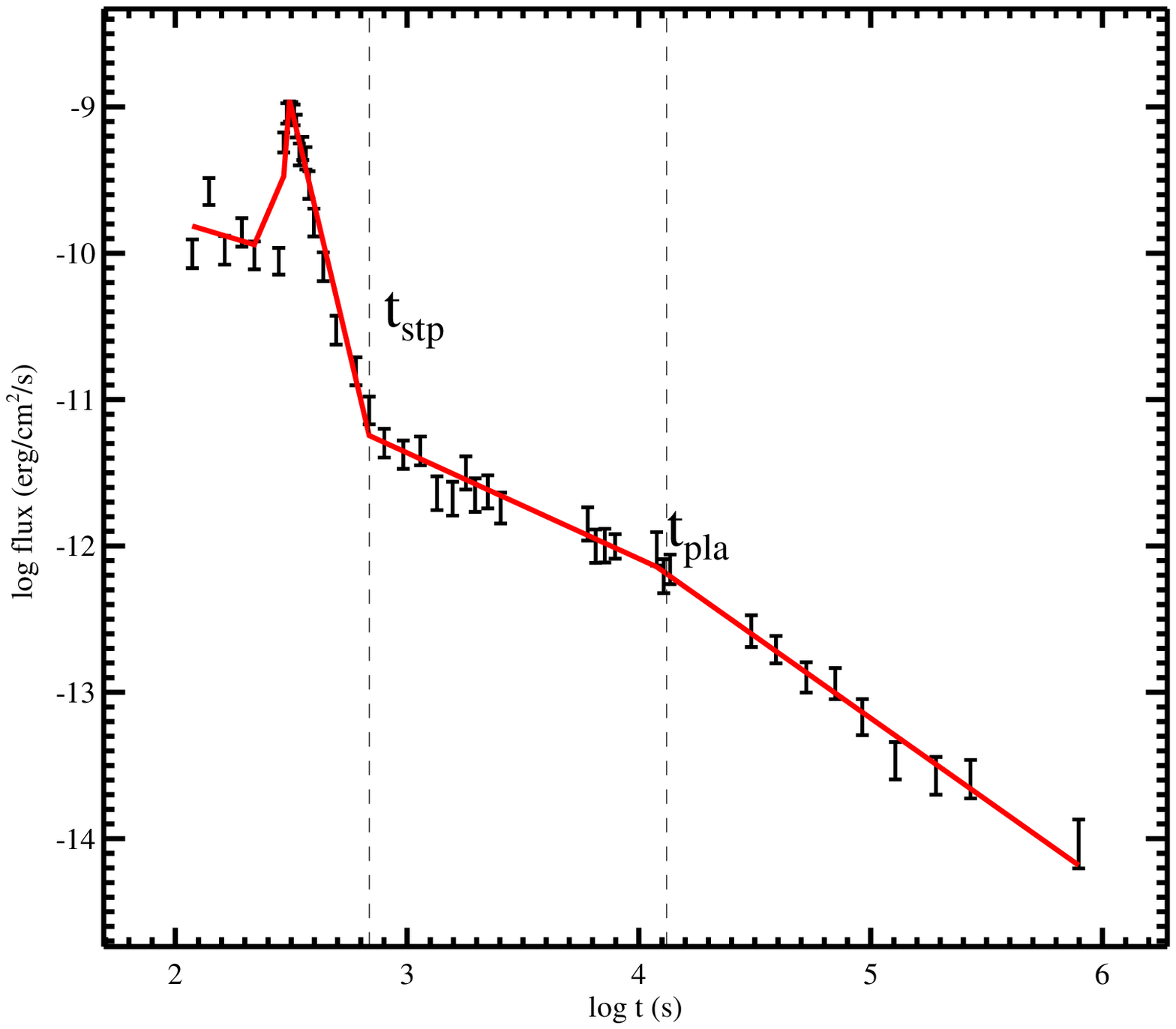}}
     \subfigure[]{
    \includegraphics[width=2.0in]{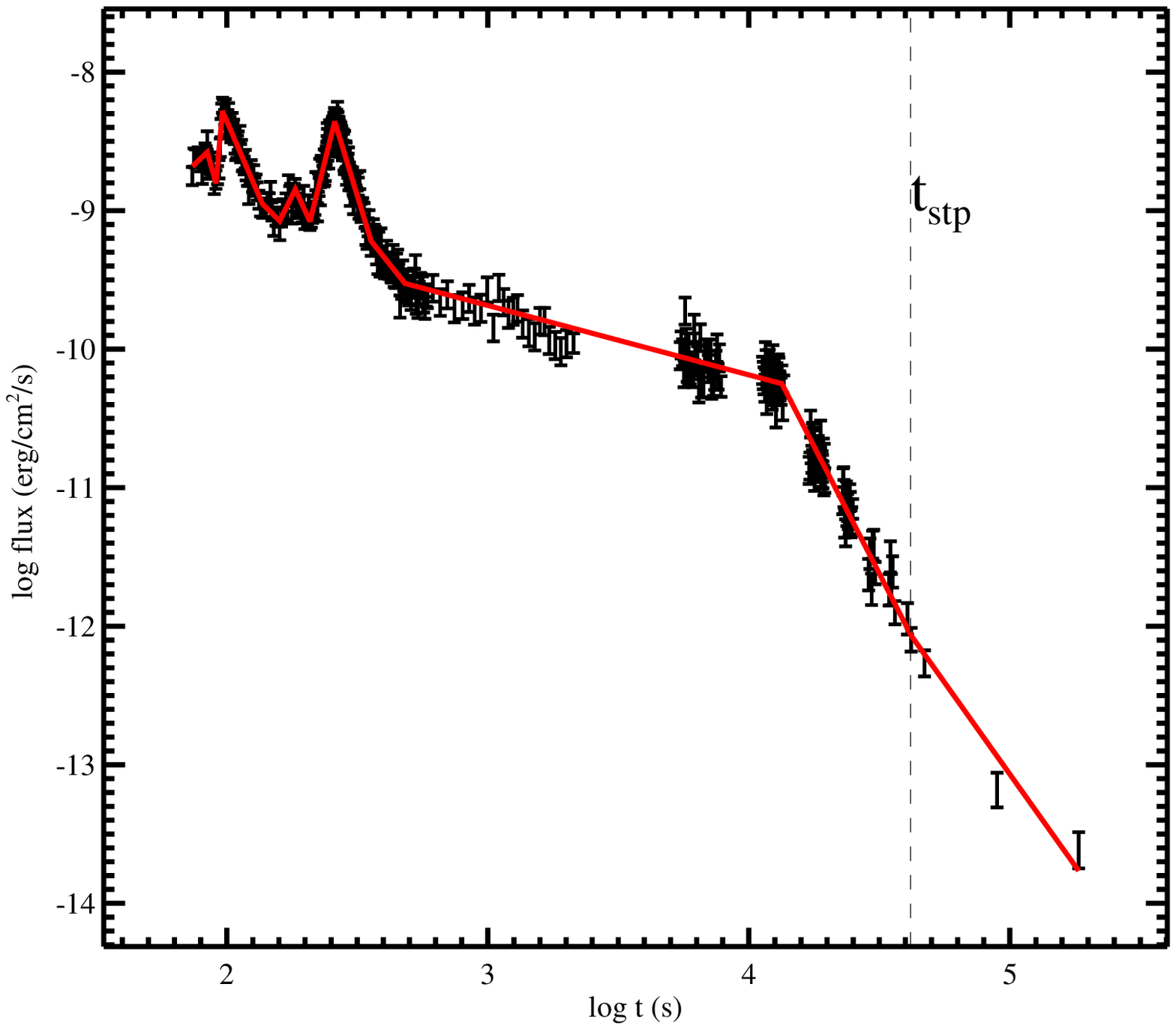}}\\
    \label{fig:examples}
    \caption{Examples of our fitting results for different types of light curves with marks of $t_{\rm stp}$ and $t_{\rm pla}$ (if applicable). The top panel (GRB 160321 and GRB 060908) is for the first category, i.e., either with simple power-law decay light curves, or with broken power-law decay light curves but without showing any steep decay or plateau signature. The first subfigure in the middle panel (GRB 151022A) is for the second category, i.e., with steep decay feature but without showing any flare or plateau feature. The other two subfigures in the middle panel (GRB 070420 and GRB 120213A) are for the third category, i.e., consisting plateau feature but without flare feature. The bottom panel is for the fourth category, i.e., consisting X-ray flare feature in the light curve.} 
            \end{figure*}

\end{document}